\pdfoutput=1

\documentclass[a4paper,11pt]{article}
\usepackage{jheppub}
\usepackage{comment}

\usepackage{hyperref}
\usepackage{cancel}

\definecolor{darkgreen}{RGB}{0,120,0}

\newcommand{\eqs}[1]{\begin{equation}\begin{split}#1\end{split}\end{equation}}

\newcommand{\eqnmy}{&=&}
\newcommand{\nonmy}{\notag \\}
\newcommand{\cD}{\mathcal{D}}
\definecolor{darkpurple}{RGB}{102, 0, 193}

\usepackage{bm, bbm, bbold}
\usepackage{latexsym}
\usepackage{dcolumn}
\usepackage{amsmath,amsfonts,amssymb, mathrsfs, amsthm}
\usepackage[T1]{fontenc}

\usepackage{graphicx,epsfig}
\usepackage{environ}
\usepackage{mathtools}
\usepackage{braket}
\usepackage{fancyhdr}
\usepackage{hyperref}
\usepackage{graphicx,epstopdf}
\usepackage{tikz}
\usepackage{float}
\usepackage{framed}
\usepackage{soul}
\usetikzlibrary{positioning,decorations.markings, decorations.pathmorphing, calc}
\tikzset{snake it/.style={decorate, decoration=snake}}
\usetikzlibrary{angles,quotes}

\usepackage{stmaryrd}
\usepackage{slashed}
\usepackage{array,multirow}

\usepackage[framemethod=default]{mdframed}
\newmdenv[skipabove=7pt,
skipbelow=7pt,
rightline=false,
leftline=false,
topline=false,
bottomline=false,
backgroundcolor=gray!10,
linecolor=gray,
innerleftmargin=5pt,
innerrightmargin=5pt,
innertopmargin=5pt,
innerbottommargin=5pt,
leftmargin=0cm,
rightmargin=0cm,
linewidth=4pt]{eBox}

\def\p{\partial}

\newcommand{\be}{\begin{equation}}
\newcommand{\ee}{\end{equation}}
\newcommand{\bes}{\begin{equation*}}
\newcommand{\ees}{\end{equation*}}

\NewEnviron{derivation}{
\begin{framed}
\begin{center}
{\bf-: Derivation :-}\\
\end{center}
  \BODY

\end{framed}
}

\NewEnviron{eqn}{
\begin{align}
\begin{split}
  \BODY
\end{split}
\end{align}
}

\NewEnviron{eqn*}{
\begin{align*}
\begin{split}
  \BODY
\end{split}
\end{align*}
}

\newcommand{\intinf}{\int_{-\infty}^{\infty}} 
\newcommand{\intsinf}{\int_{0}^{\infty}} 
\newcommand{\pd}[2]{\frac{\partial #1}{\partial #2}}

\usepackage{cancel}
\usepackage{tikz, pgf}
\usetikzlibrary{shapes.misc}
\usetikzlibrary{shapes}
\usetikzlibrary{fit}
\usetikzlibrary{arrows}

\usetikzlibrary{decorations.pathmorphing}	
\usetikzlibrary{decorations.markings}
\tikzset{
    sugra/.style={decorate, decoration={snake}, draw=black},
    scalarphi/.style={dashed,draw=black, postaction={decorate},
        },
    hwbou/.style={draw=blue, postaction={decorate}, ultra thick
        },
    vector/.style={draw=blue,decorate, decoration={snake}, draw},
	provector/.style={decorate, decoration={snake,amplitude=2.5pt}, draw},
	antivector/.style={decorate, decoration={snake,amplitude=-2.5pt}, draw},
   	 fermion/.style={draw=cyan, postaction={decorate},
        decoration={markings,mark=at position .55 with {\arrow[draw=black]{>}}}},
    fermionbar/.style={draw=cyan, postaction={decorate},
        decoration={markings,mark=at position .55 with {\arrow[draw=black]{<}}}},
    fermionnoarrow/.style={draw=black},
    gluon/.style={decorate, draw=red,
        decoration={coil,amplitude=4pt, segment length=5pt}},
    scalar/.style={dashed,draw=black, postaction={decorate},
        decoration={markings,mark=at position .55 with {\arrow[draw=black]{>}}}},
    scalarbar/.style={dashed,draw=black, postaction={decorate},
        decoration={markings,mark=at position .55 with {\arrow[draw=black]{<}}}},
    electron/.style={draw=black, postaction={decorate},
        decoration={markings,mark=at position .55 with {\arrow[draw=black]{>}}}},
    scalarnoarrow/.style={dashed, draw=black},
    electron/.style={draw=black, postaction={decorate},
        decoration={markings, mark=at position .55 with {\arrow[draw=black]{>}}}},
	bigvector/.style={decorate, decoration={snake, amplitude=4pt}, draw},
    photon/.style={draw=red, decorate, decoration={zigzag}, draw},
    higgs/.style={dashed, draw=black, postaction={decorate},
        },	
        goldstone/.style={draw=brown, postaction={decorate},
        },    
          ghost/.style={dashed, draw=magenta, postaction={decorate},
        decoration={markings, mark=at position .55 with {\arrow[draw=black]{>}}}
        },  
          antighost/.style={dashed, draw=magenta, postaction={decorate},
        decoration={markings, mark=at position .55 with {\arrow[draw=black]{<}}}
        },  
          mphoton/.style={decorate, decoration={snake}, draw=violet},
            realscalar/.style={draw=black}, 
           mgluon/.style={decorate, draw=blue,
        decoration={coil,amplitude=4pt, segment length=5pt}},
         weylfermion/.style={draw=orange, postaction={decorate},
        decoration={markings,mark=at position .55 with {\arrow[draw=black]{>}}}},
         weylfermionbar/.style={draw=orange, postaction={decorate},
        decoration={markings,mark=at position .55 with {\arrow[draw=black]{<}}}}, 
   	wboson/.style={draw=blue,decorate, decoration={snake,amplitude=4pt}, draw},  
    zboson/.style={draw=violet, decorate, decoration={snake}, draw},   
    lepton/.style={draw=black, postaction={decorate},
        decoration={markings,mark=at position .55 with {\arrow[draw=black]{>}}}},
    leptonbar/.style={draw=black, postaction={decorate},
        decoration={markings,mark=at position .55 with {\arrow[draw=black]{<}}}}, 
        graviton/.style={draw=blue,decorate, decoration={snake,amplitude=4pt}, draw},  
        gravitino/.style={draw=red, postaction={decorate},
        decoration={snake, markings, mark=at position .55 with {\arrow[draw=black]{>}}}},
    gravitinobar/.style={draw=red, postaction={decorate},
        decoration={snake, markings,mark=at position .55 with {\arrow[draw=black]{<}}} },    
}

\usepackage{subfigure}
\allowdisplaybreaks 

\title{Soft Limits of Gluon and Graviton Correlators in Anti-de Sitter Space}

\author[a]{Chandramouli Chowdhury,}
\author[b]{Arthur Lipstein,} 
\author[b]{Jiajie Mei,} 
\author[c]{Yuyu Mo} 
\affiliation[a]{Mathematical Sciences and STAG Research Centre, University of Southampton, Highfield, Southampton SO17 1BJ, United Kingdom}
\affiliation[b]{Department of Mathematical Sciences, Durham University, Stockton Road, DH1 3LE, Durham, United Kingdom}
\affiliation[c]{Higgs Centre for Theoretical Physics, School of Physics and Astronomy,
The University of Edinburgh, Edinburgh EH9 3FD, Scotland, UK}

\abstract{We derive formulae for the soft limit of tree-level gluon and graviton correlators in Anti-de Sitter space, which arise from Feynman diagrams encoding the Weinberg soft theorems in flat space. Other types of diagrams can also contribute to the soft limit at leading order in the soft momentum, but have a different pole structure. We derive these results at four points using explicit formulae recently obtained from the cosmological bootstrap and double copy, and extend them to any multiplicity using bootstrap techniques in Mellin-momentum space.}

\begin{document}

\maketitle

\noindent
\flushbottom
\allowdisplaybreaks

\section{Introduction}

A major driving force in the study of scattering amplitudes was the discovery of soft theorems which relate higher point amplitudes to lower point amplitudes when the momentum of one external leg is taken to zero. Soft gluon and graviton theorems were first derived at leading order in the soft momentum by Weinberg \cite{Weinberg:1965nx} and then extended to higher orders in \cite{Low:1958sn,Burnett:1967km,White:2011yy,Cachazo:2014fwa,Casali:2014xpa,Hui:2018cag}. They are of great interest because they imply useful constraints on scattering amplitudes and encode hidden symmetries. For example in certain scalar theories they encode hidden shift symmetries via enhanced soft limits \cite{Adler:1964um,Arkani-Hamed:2008owk,Cheung:2016drk,Cheung:2014dqa,Hinterbichler:2015pqa},  while in Yang-Mills (YM) and Einstein gravity they encode asymptotic symmetries \cite{Strominger:2013jfa,He:2014laa,Strominger:2017zoo} known as extended Bondi-Metzner-Sachs (BMS) symmetry \cite{Bondi:1962px,Sachs:1962wk}. 

Soft limits also play an important role in cosmology where they give rise to consistency conditions on inflationary correlators \cite{Maldacena:2002vr,Creminelli:2004yq,Cheung:2007sv,Creminelli:2012ed,Hinterbichler:2012nm,Bzowski:2012ih,Hinterbichler:2013dpa,Pimentel:2013gza,McFadden:2014nta} and allow one to deduce certain inflationary three-point functions from soft limits of four-point de Sitter (dS) correlators \cite{Kundu:2014gxa,Creminelli:2003iq,Assassi:2012zq,Kundu:2015xta}. More recently, they were also shown to encode hidden symmetries of certain scalar theories via enhanced soft limits analogous to those in flat space \cite{Bonifacio:2021mrf,Armstrong:2022vgl}. The soft limit of three-point graviton correlators in de Sitter was found long ago by Maldacena \cite{Maldacena:2002vr} and takes the form of an energy derivative acting on a two-point correlators. Similar structure was also found for inflationary correlators with a soft graviton and any number of scalars using symmetry-based arguments\cite{Creminelli:2012ed,Hinterbichler:2013dpa}. Our goal in this paper will be to derive a formula for the soft limits of gluon and graviton correlators with any multiplicity in Anti-de Sitter space (AdS). Curvature corrections to the Weinberg soft theorems in AdS were found by first expanding around the flat space limit \cite{Banerjee:2021llh,AtulBhatkar:2021sdr,Banerjee:2022oll,Fernandes:2023xim}, but in this paper we will not perform such an expansion. In fact, we will show that the soft and flat space limits do not commute.

First we compute the soft limits of the tree-level four-point gluon and graviton correlators in AdS$_4$. While this is fairly straightforward to do for gluons using Feynman diagrams, for gravitons we make use of a compact formula for the four-point graviton correlator recently derived using a combination of bootstrap \cite{Bonifacio:2022vwa} and double copy \cite{Armstrong:2023phb} techniques. We then obtain soft limit formulae for arbitrary multiplicity by computing the soft limit of class I diagrams which give rise to the Weinberg soft theorems in flat space. Unlike in flat space, other types of diagrams can also contribute to the soft limit at leading order in the soft momentum, but they have a different pole structure than class I diagrams so they can be distinguished, as we will explain. The soft limit of class I diagrams is very difficult to prove in gravity using standard Feynman diagram methods, so we prove it using bootstrap techniques in Mellin-momentum space \cite{Sleight:2019hfp,Sleight:2019mgd,Mei:2023jkb,Mei:2024abu}. This approach makes use of a differential representation where correlators are represented in terms of certain differential operators acting on bulk scalar contact diagrams \cite{Roehrig:2020kck,Eberhardt:2020ewh,Gomez:2021qfd,Herderschee:2022ntr}.

Ultimately, we find that the soft limit of gluon and graviton correlators gives a term analogous to the Weinberg soft factor for scattering amplitudes where the soft pole is replaced with an energy derivative plus a term involving a polarisation derivative which is subleading in the flat space limit. These two terms can be nicely combined into a single momentum derivative acting on the bulk-to-boundary propagator of a hard leg. As mentioned above, similar structures appear in the consistency relations for inflationary correlators, which are derived from Ward identities associated with certain nonlinearly realised large diffeomeorphisms \cite{Creminelli:2012ed,Hinterbichler:2013dpa}. On the other hand, we derive soft limits diagramatically rather than from underlying symmetry principles. It would be very interesting to see how the soft limits in this paper are related to those of inflationary correlators or to derive them from the analogoue of BMS symmetry in AdS known as $\Lambda$-BMS symmetry, which was recently discovered in \cite{Compere:2019bua}.  

The structure of this paper is as follows. In section \ref{review} we review the Weinberg soft theorems in flat space and the soft limit of three-point gluon and graviton correlators in AdS$_4$. In section \ref{4point} we compute the soft limits of four-point gluon and graviton correlators in AdS$_4$ and in section \ref{sec:generalMult} we obtain soft limit formulae for any multiplicity by computing the soft limit of general class I diagrams using bootstrap techniques in Mellin-momentum space. In section \ref{conclusion}, we present our conclusions. There are also several of Appendices which describe various technical details. For example, we analyze soft limits of general YM diagrams using Feynman rules in AdS momentum space and show that non-class I diagrams exhibit different pole structure than class I diagrams. We also verify our soft limit formula for five-point correlators in YM. We also  include two Mathematica notebooks, \texttt{Class1.nb} and \texttt{5ptYM.nb}, which provide details of the bootstrap procedure for class I diagrams and our five-point checks, respectively.  

\section{Review} \label{review}

As shown long ago by Weinberg \cite{Weinberg:1965nx}, soft limits of gluon and graviton ampitudes have the following universal form:
\begin{align}
\lim_{k_{n+1}^{\mu}\rightarrow0}\mathcal{\mathcal{A}}^{YM}_{n+1}&=\left(\sum_{a=1}^n\frac{\varepsilon_{n+1}\cdot k_{a}}{k_{n+1}\cdot k_{a}}+...\right)\mathcal{\mathcal{A}}^{YM}_{n}\\
\lim_{k_{n+1}^{\mu}\rightarrow0}\mathcal{A}^{GR}_{n+1}&=\left(\sum_{a=1}^{n}\frac{\left(\varepsilon_{n+1}\cdot k_{a}\right)^{2}}{k_{n+1}\cdot k_{a}}+...\right)\mathcal{A}^{GR}_{n}
\label{weinbergsoft}
\end{align}
where ... denote subleading terms in the soft expansion. In recent years, deep connections to asympotic symmetries have been identified \cite{Strominger:2013jfa,He:2014laa}.  In this paper our goal will be to find analogous formulae for boundary correlators in AdS. Before doing so, we will first review some facts about AdS correlators and soft limits of 3-point functions. 

We will work in the Poincare patch of Euclidean AdS$_{d+1}$ with unit radius:
\begin{equation}
	ds^{2}=\frac{dz^{2}+dx^{i}dx^{i}}{z^{2}},
\end{equation}
where $0<z<\infty$ is the radial coordinate and $x^i$ with $i\in\left\{ 1,...,d\right\} $ are the coordinates of the boundary located at $z=0$. In AdS, the basic obervables are computed by summing over Feynman diagrams ending on the boundary, which can be formally treated like boundary CFT correlators. Wick-rotating to de Sitter space then gives cosmological wavefunction coefficients \cite{Maldacena:2002vr,McFadden:2009fg}. In-in correlators can then be obtained by squaring the wavefunction and performing a path integral over boundary values of the bulk fields \cite{Maldacena:2002vr,Weinberg:2005vy}, although we will not consider such objects in this paper. Since the boundary is translation invariant, we Fourier transform correlators along the boundary to momentum space to get \cite{Bzowski:2013sza}
\begin{equation}
\left\langle \mathcal{O}_{\Delta}\left(\vec{k}_{1}\right)...\mathcal{O}_{\Delta}\left(\vec{k}_{n}\right)\right\rangle =\delta^{d}\left(\vec{k}_{T}\right)\left\langle \left\langle \mathcal{O}_{\Delta}\left(\vec{k}_{1}\right)...\mathcal{O}_{\Delta}\left(\vec{k}_{n}\right)\right\rangle \right\rangle,
\end{equation}
where $\vec{k}_T$ is the total boundary momentum, which must vanish by momentum conservation:
\begin{equation}
\vec{k}_T=\sum_{a=1}^{n}\vec{k}_{a}=0,
\end{equation}
where the subscript $a$ labels external legs and we denote the correlator stripped of the delta function with double brackets. For simplicity, we consider scalar operators in the boundary with scaling dimension $\Delta$ which are dual to bulk scalar fields $\phi$ with mass $m^2=\Delta\left(\Delta-d\right)$. The bulk-to-boundary propagators for the bulk scalar fields are obtained by solving the free equations of motion \cite{Liu:1998ty,Raju:2011mp}
\begin{equation}
\mathcal{D}_{k}^{\Delta}\phi_{\Delta}(k,z)=0,\,\,\,\mathcal{D}_{k}^{\Delta}\equiv z^{2}k^{2}-z^{2}\partial_{z}^{2}-(1-d)z\partial_{z}+\Delta(\Delta-d),
\label{scalareom}
\end{equation}
where $k=|\vec{k}|$ is the norm of the boundary momentum flowing through the propagator, which we refer to as the energy of the particle\footnote{More precisely, this can be thought of as the radial component of the momentum in AdS, but after Wick rotation to dS the radial direction becomes time-like so this can be interpreted as an energy.}. In general, the total energy of an $n$-point correlator
\begin{equation}
k_{12...n}=\sum_{a=1}^{n}k_{a},\,\,\,k_{a}=\left|\vec{k}_{a}\right|
\end{equation}
is not conserved. In the flat space limit, the energy is taken to zero and the correlator develops a pole whose residue is the flat space amplitude in $(d+1)$ dimensions \cite{Raju:2012zr}. 

In more detail, the solution to \eqref{scalareom} is given by
\begin{equation}
\phi_{\Delta}(k, z)=\sqrt{\frac{2}{\pi }} z^{d/2} k^{\Delta -\frac{d}{2}} K_{\Delta -\frac{d}{2}}(z	k)\label{Btb}
\end{equation}
where $K$ is a Bessel function. A useful fact is that the bulk-to-boundary propagators for gluons and gravitons can be obtained by dressing scalar propagators with polarisations \cite{Liu:1998ty,Raju:2011mp} 
\begin{equation}
A_{i}\left(k,z\right)=\varepsilon_{i}\phi_{\Delta=d-1}\left(k,z\right),\,\,\,h_{ij}=\varepsilon_{ij}\phi_{\Delta=d}\left(k,z\right),
\label{spinningprops}
\end{equation}
where we have rescaled the propagators of by factors of $z$ and absorbed these factors into the interaction vertices (for more details about this and the gauge choice see \cite{Armstrong:2022mfr}). Polarisations point along the boundary and satisfy
\begin{equation}
\vec{\varepsilon}_{a}\cdot\vec{k}_{a}=\vec{\varepsilon}_{a}\cdot\vec{\varepsilon}_{a}=0.
\label{polarisationconstraint}
\end{equation}
Moreover, graviton polarisations can be written as a tensor product of gluon polarisations:
\begin{equation}
\varepsilon_{a}^{ij}=\varepsilon_{a}^{i}\varepsilon_{a}^{j}.
\label{gravpol}
\end{equation}
We will not go into further details about the Feynman rules for gluons and gravitons in AdS since later on we will make use of bootstrap techniques that will not require a detailed knowledge of the Feynman rules.

Correlators of gluons and gravitons can be represented as boundary correlators of conserved currents $J$ and stress tensors $T$, respectively. The two and three-point gluon correlators in AdS$_{4}$ are given by \cite{Maldacena:2011nz,Farrow:2018yni}
\begin{equation}
\langle \langle JJ \rangle \rangle=-\frac{1}{2}k_{2}\varepsilon_{1}\cdot\varepsilon_{2},\,\,\,\langle \langle JJJ \rangle \rangle=\frac{1}{k_{123}}\left(\varepsilon_{1}\cdot\varepsilon_{2}\varepsilon_{3}\cdot k_{1}+{\rm cyclic}\right),
\label{ym2pt3pt}
\end{equation}
where the double brackets once again indicate that we drop the momentum-conserving delta function \footnote{We have chosen the normalisation of two point functions for consistency with soft limits.
}. If we take the soft limit of the three-point we obtain
\begin{equation}
\lim_{\vec{k}_{3}\rightarrow0}\langle \langle JJJ \rangle \rangle=-\frac{\varepsilon_{3}\cdot k_{2}\varepsilon_{1}\cdot\varepsilon_{2}}{2k_{2}}=-\frac{\varepsilon_{3}\cdot k_{2}}{2k_{2}}\partial_{k_{2}}\langle \langle JJ \rangle \rangle,
\label{3ptym}
\end{equation}
where the energy derivative acts trivially on polarisations so only acts on the $k_2$ in \eqref{ym2pt3pt}. Similarly, the two and three-point graviton correlators in AdS$_4$ are given by \cite{Maldacena:2011nz,Farrow:2018yni}:
\begin{eqnarray}
\langle \langle TT \rangle \rangle&=&\frac{1}{2}k_{2}^{3}\left(\varepsilon_{1}\cdot\varepsilon_{2}\right)^{2},\\
\langle \langle TTT \rangle \rangle&=&\left(\frac{k_{1}k_{2}k_{3}}{k_{123}^{2}}+\frac{k_{1}k_{2}+k_{2}k_{3}+k_{3}k_{1}}{k_{123}}-k_{123}\right)\left(\varepsilon_{1}\cdot\varepsilon_{2}\varepsilon_{3}\cdot k_{1}+{\rm cyclic}\right)^{2},
\label{2pt3ptgr}
\end{eqnarray}
where we neglect boundary contact terms in the three-point expression
which can be removed by redefining the bulk metric. For simplicity, we also neglect a well-known IR divergent term \cite{Maldacena:2011nz} which cancels out of the in-in correlator and will not effect our results later on. Taking the soft limit of the three-point correlator then gives:
\begin{equation}
\lim_{\vec{k}_{3}\rightarrow0}\langle \langle TTT \rangle \rangle=-\frac{3}{2}k_{2}\left(\varepsilon_{1}\cdot\varepsilon_{2}\varepsilon_{3}\cdot k_{2}\right)^{2}=-\frac{\left(\varepsilon_{3}\cdot k_{2}\right)^{2}}{2k_{2}}\partial_{k_{2}}\langle \langle TT \rangle \rangle.
\label{3ptgr}
\end{equation}
A similar formula for in-in correlators was first found by Maldacena in the context of cosmology \cite{Maldacena:2002vr}. Note that the divergent term we neglected in \eqref{2pt3ptgr} does not contribute to the soft limit because the terms which survive in the soft limit are analytic in at least two momenta and therefore correspond to boundary contact terms at three points \cite{Maldacena:2011nz}. 

Our goal in this paper will be to generalise these formulae to arbitrary multiplicity. For the impatient reader, our main results can be found in \eqref{wqkjhf} and \eqref{grsoftresmom}. As we will see later, the energy derivatives in \eqref{3ptym} and \eqref{3ptgr} can be thought of as the AdS analogue of Weinberg soft factors in \eqref{weinbergsoft}, and appear at arbitrary multplicity. Above three-points, we also obtain polarisation derivatives which are subleading in the flat space limit. Similar structures also appear in the soft limits of inflationary correlators \cite{Hinterbichler:2013dpa,Creminelli:2012ed} so it would be very interesting to understand how they are related to the soft limits of boundary correlators in AdS. Note that the relation is not direct because cosmological correlators are obtained by squaring the cosmological wavefunction, which maps onto boundary correlators in AdS after analytic continuation \cite{Maldacena:2002vr,Weinberg:2005vy}. 

\section{Four Points} \label{4point}

In this section we compute the soft limit of four-point gluon and graviton correlators in AdS$_4$ using explicit four-point correlators recently derived in \cite{Armstrong:2023phb} using double copy techniques. We will observe a Weinberg-like soft factor involving an energy derivative as well as another term with a polarisation derivative which is subleading in the flat space limit. As we will show in the next section, this structure extends to any number of points.

\subsection{Yang-Mills} \label{ym4ptsection}

Let us consider the color-ordered tree-level four-point correlator. This is given by a sum of s and t-channel diagrams along with a contact diagram as depicted in the figure below:
\begin{eqn}
\begin{tikzpicture}[baseline]

\draw[very thick, black] (-1, 0) -- (1,0);
\draw[very thick, black] (1, 0) -- ({2.25*cos(20)}, {2.25*sin(20)});
\draw[very thick, black] (1, 0) -- ({2.25*cos(-20)}, {2.25*sin(-20)});
\draw[very thick] (0, 0) circle (2.25);

\draw[very thick, black] (-1, 0) -- ({2.25*cos(160)}, {2.25*sin(160)});
\draw[very thick, black] (-1, 0) -- ({2.25*cos(200)}, {2.25*sin(200)});



\node at ({2.5*cos(160)}, {2.5*sin(160)}) {$2$};
\node at ({2.5*cos(20)}, {2.5*sin(20)}) {$3$};
\node at ({2.5*cos(-20)}, {2.5*sin(-20)}) {$4$};
\node at ({2.5*cos(200)}, {2.5*sin(200)}) {$1$};

\end{tikzpicture} 
 \quad  + \quad (2 \leftrightarrow 4) \quad + 
 \quad 
 \begin{tikzpicture}[baseline]

\draw[very thick, black] (0, 0) -- ({2.25*cos(45)}, {2.25*sin(45)});
\draw[very thick, black] (0, 0) -- ({2.25*cos(-45)}, {2.25*sin(-45)});
\draw[very thick] (0, 0) circle (2.25);

\draw[very thick, black] (0, 0) -- ({2.25*cos(135)}, {2.25*sin(135)});
\draw[very thick, black] (0, 0) -- ({2.25*cos(-135)}, {2.25*sin(-135)});



\node at ({2.5*cos(135)}, {2.5*sin(135)}) {$2$};
\node at ({2.5*cos(45)}, {2.5*sin(45)}) {$3$};
\node at ({2.5*cos(-45)}, {2.5*sin(-45)}) {$4$};
\node at ({2.5*cos(-135)}, {2.5*sin(-135)}) {$1$};

\end{tikzpicture} 
\end{eqn}
It is possible to express this sum over Feynman diagrams (including the contact diagram) as a sum of s and t-channel contributions. Using Feynman rules in momentum space, it is not difficult to show that the s-channel contribution is \cite{Armstrong:2023phb}
\begin{equation}
\langle \langle JJJJ \rangle \rangle^{(s)}=\frac{W_{s}}{k_{1234}E_{L}E_{R}}+\frac{\varepsilon_{1}\cdot\varepsilon_{2}\varepsilon_{3}\cdot\varepsilon_{4}\Pi_{1,1}}{k_{1234}E_{L}E_{R}}-\frac{\varepsilon_{1}\cdot\varepsilon_{2}\varepsilon_{3}\cdot\varepsilon_{4}\Pi_{1,0}}{k_{1234}
}+\frac{V_{c}^s}{4k_{1234}},
\label{4ptgluon}
\end{equation}
where $E_L=k_{12}+k_S$, $E_R=k_{34}+k_S$, $k_{S}=\left|\vec{k}_{12}\right|$,
	\begin{align}
	W_s=	& \varepsilon_1 \cdot \varepsilon_2\left(k_1 \cdot \varepsilon_3 k_2 \cdot \varepsilon_4-k_2 \cdot \varepsilon_3 k_1 \cdot \varepsilon_4\right) \\
		& +\varepsilon_3 \cdot \varepsilon_4\left(k_3 \cdot \varepsilon_1 k_4 \cdot \varepsilon_2-k_4 \cdot \varepsilon_1 k_3 \cdot \varepsilon_2\right) \\
		& +\left(k_2 \cdot \varepsilon_1 \varepsilon_2-k_1 \cdot \varepsilon_2 \varepsilon_1\right) \cdot\left(k_4 \cdot \varepsilon_3 \varepsilon_4-k_3 \cdot \varepsilon_4 \varepsilon_3\right)\\
\Pi_{1,1}&=\frac{1}{4}\left(k_1-k_2\right) \cdot\left(k_3-k_4\right)+\frac{\left(k_1^2-k_2^2\right)\left(k_3^2-k_4^2\right)}{4 k_S^2}\\
\Pi_{1,0}&=\frac{\left(k_1-k_2\right)\left(k_3-k_4\right)}{4 k_S^2}\\
 V_c^s&=(\varepsilon_1 \cdot \varepsilon_3 \varepsilon_2 \cdot \varepsilon_4-\varepsilon_1 \cdot \varepsilon_4 \varepsilon_2 \cdot \varepsilon_3).
\label{definitions}
	\end{align}
If we take the soft limit of $\vec{k}_{4}$, we obtain
\begin{equation}
\lim_{\vec{k}_{4}\rightarrow0}\langle \langle JJJJ \rangle \rangle^{(s)}=-\frac{\varepsilon_{4}\cdot k_{3}}{2k_{3}k_{123}}\langle \langle JJJ \rangle \rangle-\frac{\varepsilon_{1}\cdot\varepsilon_{2}\varepsilon_{3}\cdot\varepsilon_{4}\left(k_{1}-k_{2}\right)}{4k_{3}k_{123}}+\frac{V_{c}^{(s)}}{4k_{123}},\label{YM4soft}
\end{equation}
where we noted that 
\begin{equation}
\lim_{\vec{k}_{4}\rightarrow0}\Pi_{1,1}=0,\,\,\,\lim_{\vec{k}_{4}\rightarrow0}\Pi_{1,0}=\frac{k_{1}-k_{2}}{4k_{3}}.
\end{equation}

To proceed, we drop the contribution from the contact digram $V_{c}^{(s)}$ in \eqref{YM4soft} and re-write the remaining expression as
\begin{equation}
\lim_{\vec{k}_{4}\rightarrow0}\langle \langle JJJJ \rangle \rangle^{(s)}=\frac{\varepsilon_{4}\cdot k_{3}}{2k_{3}}\partial_{k_{3}}\langle \langle JJJ \rangle \rangle+\frac{\varepsilon_{3}\cdot\varepsilon_{4}}{2k_{3}^{2}}k_{3}^{i}V_{12}^{i}\int dze^{-k_{12}z}\left(1-e^{-k_{3}z}\right)+...\label{eq:ym4ptto3pt}
\end{equation}
where the energy derivative only acts on the energy pole of the three-point correlator in \eqref{ym2pt3pt},
\begin{equation}
V_{ab}^{i}=\frac{1}{2}\varepsilon_{a}\cdot\varepsilon_{b}\left(k_{a}-k_{b}\right)^{i}+\varepsilon_{a}\cdot k_{b}\varepsilon_{b}^{i}-\varepsilon_{b}\cdot k_{a}\varepsilon_{a}^{i}
\end{equation}
is the three-point gluon vertex, and ... denote the terms we have dropped. As we will see in the next section, the terms we have kept arise from the soft limit of Feynman diagrams which encode the Weinberg soft limit in flat space (known as class I diagrams) and the terms we have dropped arise from other classes of diagrams. To obtain \eqref{eq:ym4ptto3pt} we noted that $k_{3}^{i}V_{12}^{i}=-\frac{1}{2}\varepsilon_{1}\cdot\varepsilon_{2}\left(k_{1}^{2}-k_{2}^{2}\right)$ 
and 
\begin{equation}
\int dze^{-k_{12}z}\left(1-e^{k_{3}z}\right)=\frac{k_{3}}{k_{123}k_{12}}.
\end{equation}
We may then write \eqref{eq:ym4ptto3pt} as follows: 
\begin{equation}
\lim_{\vec{k}_{4}\rightarrow0}\langle \langle JJJJ \rangle \rangle^{(s)}=\frac{\varepsilon_{4}\cdot k_{3}}{2k_{3}}\partial_{k_{3}}\langle \langle JJJ \rangle \rangle+\frac{\varepsilon_{3}\cdot\varepsilon_{4}}{2k_{3}^{2}}k_{3}\cdot\partial_{\varepsilon_{3}}\left(\left.\langle \langle JJJ \rangle \rangle\right|_{k_{3}=0}-\langle \langle JJJ \rangle \rangle\right)+...\label{eq:YMsoftOnshell4to3}
\end{equation}
where $\left.\langle \langle JJJ \rangle \rangle\right|_{k_{3}=0}$ means that the energy $k_3$ is set to zero but not $\vec{k}_3$, so this restriction just means that we take the energy pole $k_{123}^{-1}\rightarrow k_{12}^{-1}$. Note that the first term in parenthesis in \eqref{eq:YMsoftOnshell4to3} corresponds to a boundary contact term in position space and can be dropped. To see this, let's write out this term explicitly:
\begin{equation}
\frac{\varepsilon_{3}\cdot\varepsilon_{4}}{2k_{3}^{2}}k_{3}\cdot\varepsilon_{3}\left.\langle \langle JJJ \rangle \rangle\right|_{k_{3}=0}=\frac{\varepsilon_{3}\cdot\varepsilon_{4}}{2k_{3}^{2}}k_{3}^{i}V_{12}^{i}\int dze^{-k_{12}z}=\frac{\varepsilon_{1}\cdot\varepsilon_{2}\varepsilon_{3}\cdot\varepsilon_{4}}{4k_{3}^{2}}\left(k_{1}-k_{2}\right)
\end{equation}
The first term on the right-hand-side is analytic in legs 2 and 3 while the second term is analytic in legs 1 and 3. As shown in \cite{Maldacena:2011nz}, terms which are analytic in at least two momenta give delta functions after Fourier transforming to position space so vanish for generic positions of the dual operators. 

In summary, we find the following soft limit for the s-channel contribution to a 4-point gluon correlator:
\begin{align}
\lim_{\vec{k}_{4}\rightarrow0}\langle \langle JJJJ \rangle \rangle^{(s)} &=\frac{\varepsilon_{4}\cdot k_{3}}{2k_{3}}\partial_{k_{3}}\langle \langle JJJ \rangle \rangle-\frac{\varepsilon_{4}\cdot\varepsilon_{3}}{2k_{3}^{2}}k_{3}\cdot\partial_{\varepsilon_{3}}\langle \langle JJJ \rangle \rangle+... \label{4ptsoft1} \\
&=\frac{1}{2}\varepsilon_{4}^{i}\partial_{k_{3i}}\langle\langle JJJ\rangle\rangle+...,
\label{4ptsoft}
\end{align}
where the momentum derivative in the second line acts on the bulk-to-boundary propagator of leg 3. Recalling that the bulk-to-boundary propagator can be expressed in terms of scalar propagator (which only depends on energy) dressed with a polarisation, we see that the momentum derivative in \eqref{4ptsoft} can act on the energy or polarisation of the propagator, which gives \eqref{4ptsoft1} after using the following identities:
\begin{equation}
\partial_{k^{i}}=\frac{k_{i}}{k}\partial_{k},\,\,\,\frac{\partial\varepsilon_{i}}{\partial k^{j}}=-\frac{\varepsilon_{j}k_{i}}{k^{2}}.
\label{momdervident}
\end{equation}
The first one follows from the chain rule while the second one follows from taking a derivative of \eqref{polarisationconstraint}. The soft limit of the full color-ordered gluon correlator is then obtained by summing over the s and t channels. 

Let us comment on the physical interpretation of \eqref{4ptsoft}. Consider the first term on the right hand side of \eqref{4ptgluon}, which comes from gluon exchange. Notice that it has a simple pole in the total energy $k_{1234}$. Taking the energy to zero gives
\begin{equation}
\lim_{k_{1234}\rightarrow0}\frac{W_{s}}{k_{1234}E_{L}E_{R}}=\frac{1}{k_{1234}}\frac{W_{s}}{S},
\end{equation}
where $S=k_{34\mu}k_{34}^{\mu}$ is the 4d Lorentz-invariant Mandelstam variable. We recognise the residue of the energy pole to be the s-channel gluon exchange diagram in flat space. If we then take the soft limit $\vec{k}_{4}\rightarrow0$ we get
\begin{equation}
\lim_{\vec{k}_{4}\rightarrow0}\lim_{k_{1234}\rightarrow0}\frac{W_{s}}{k_{1234}E_{L}E_{R}}=\frac{1}{k_{123}}\frac{\varepsilon_{4}\cdot k_{3}}{k_{4}\cdot k_{3}}\left(\varepsilon_{1}\cdot\varepsilon_{2}\varepsilon_{3}\cdot k_{1}+{\rm cyclic}\right),
\end{equation}
which we recognize as a term contributing to the Weinberg soft gluon theorem in \eqref{weinbergsoft} times a three-point energy pole. If we instead take the soft limit followed by the flat space limit, we obtain
\begin{equation}
\lim_{k_{1234}\rightarrow0}\lim_{\vec{k}_{4}\rightarrow0}\frac{W_{s}}{k_{1234}E_{L}E_{R}}=\frac{\varepsilon_{4}\cdot k_{3}}{2k_{3}k_{123}^{2}}\left(\varepsilon_{1}\cdot\varepsilon_{2}\varepsilon_{3}\cdot k_{1}+{\rm cyclic}\right).
\end{equation}
Hence, the soft and flat space limits do not commute and we get a double pole in $k_{123}$ which arises from acting with $\partial_{k_3}$ on the energy pole of the three-point function in \eqref{4ptsoft}. The double pole can in turn be written as a single pole in $k_{123}$ times a pole in the energy of the soft leg $k_4$ using energy conservation associated with the flat space limit. Hence, we see that the first term on the right-hand-side of \eqref{4ptsoft1} is indeed the analogue of the Weinberg soft factor in AdS. In addition to this, we also find a polarisation derivative in \eqref{4ptsoft1} which is subleading in the flat space limit. The two terms combine nicely into a single derivative with respect to boundary momentum acting on the bulk-to-boundary propagator of leg 3.

\subsection{Gravity}

Next let us consider the soft limit of the 4-point graviton correlator. A compact
formulae was obtained from
the double copy in \cite{Armstrong:2023phb}. It can
be written as a sum over three channels and we give the explicit formula
for the s-channel and compute its soft limit in Appendix \ref{4ptgr}.
In the end we obtain 
\begin{align}
\lim_{\vec{k}_{4}\rightarrow0}\langle \langle TTTT \rangle \rangle^{(s)}&={\color{black}\frac{1}{2}}\left(k_{3}\cdot\varepsilon_{4}\right)^{2}\left(\frac{2k_{1}k_{2}}{k_{123}^{3}}+\frac{k_{12}}{k_{123}^{2}}+\frac{1}{k_{123}}\right)\left(\varepsilon_{1}\cdot\varepsilon_{2}\varepsilon_{3}\cdot k_{1}+{\rm cyclic}\right)^{2}\nonmy
& +{\color{black}\frac{1}{2}}k_{3}\cdot\varepsilon_{4}\left(\varepsilon_{1}\cdot\varepsilon_{2}\varepsilon_{3}\cdot k_{1}+{\rm cyclic}\right)\varepsilon_{1}\cdot\varepsilon_{2}\varepsilon_{3}\cdot\varepsilon_{4}\left(k_{1}-k_{2}\right)\left(\frac{k_{1}k_{2}}{k_{123}^{2}}+\frac{k_{12}}{k_{123}}\right)\nonmy
&{\color{black} -\frac{1}{2}}k_{3}\cdot \varepsilon_{4}\left(\varepsilon_{1}\cdot\varepsilon_{2}\varepsilon_{3}\cdot k_{1}+{\rm cyclic}\right)V_{c}^{(s)}\left(\frac{k_{1}k_{2}k_{3}}{k_{123}^{2}}+\frac{k_{1}k_{2}+k_{2}k_{3}+k_{3}k_{1}}{k_{123}}-k_{123}\right).
\label{4ptgrsoft1}
\end{align}
We see that the soft limit can be written in terms of gluonic
building blocks, reflecting the underlying double copy structure. As we did in the previous subsection, we drop the gluonic four-point vertex $V_{c}^{(s)}$ and re-write the remaining terms as follows:
\begin{equation}
\lim_{\vec{k}_{4}\rightarrow0}\langle \langle TTTT \rangle \rangle^{(s)}=-\frac{\left(\varepsilon_{4}\cdot k_{3}\right)^{2}}{2k_{3}}\partial_{k_{3}}\langle \langle TTT \rangle \rangle-\frac{\varepsilon_{4}\cdot k_{3}\varepsilon_{3}\cdot\varepsilon_{4}}{2k_{3}^{2}}\varepsilon_{3}^{(i}k_{3}^{j)}\tilde{V}_{12}^{ij}\frac{k_{3}^{2}}{k_{12}}\left(\frac{k_{1}k_{2}}{k_{123}^{2}}+\frac{k_{12}}{k_{123}}\right)+...,\label{eq:gr4ptto3pt}
\end{equation}
where ... denote the terms we have dropped, $\varepsilon_{3}^{(i}k_{3}^{j)}\equiv \varepsilon_{3}^{i}k_{3}^{j}+\varepsilon_{3}^{j}k_{3}^{i}$, $\tilde{V}_{ab}^{ij}=V_{ab}^{i}V_{ab}^{j}$,
and we noted that
\begin{equation}
-\frac{1}{k_{3}}\frac{\partial}{\partial k_{3}}\left(\frac{k_{1}k_{2}k_{3}}{k_{123}^{2}}+\frac{k_{1}k_{2}+k_{2}k_{3}+k_{3}k_{1}}{k_{123}}-k_{123}\right)=\frac{2k_{1}k_{2}}{k_{123}^{3}}+\frac{k_{12}}{k_{123}^{2}}+\frac{1}{k_{123}},
\end{equation}
\begin{equation}
\varepsilon_{3}^{(i}k_{3}^{j)}\tilde{V}_{12}^{ij}=-\frac{1}{2}\left(\varepsilon_{1}\cdot\varepsilon_{2}\varepsilon_{3}\cdot k_{1}+{\rm cyclic}\right)\varepsilon_{1}\cdot\varepsilon_{2}\left(k_{1}^{2}-k_{2}^{2}\right).
\end{equation}
As we will see in the next section, \eqref{eq:gr4ptto3pt} encodes the contribution of Feynman diagrams which encode the Weinberg soft theorem in flat space (known as class I diagrams) and the terms we have neglected arise from other types of diagrams. We can recast the gravitational soft limit in \eqref{eq:gr4ptto3pt}
in a form analogous to the gluonic one in \eqref{eq:ym4ptto3pt}: 
\begin{equation}
\lim_{\vec{k}_{4}\rightarrow0}\langle \langle TTTT \rangle \rangle^{(s)}=-\frac{\left(\varepsilon_{4}\cdot k_{3}\right)^{2}}{2k_{3}}\partial_{k_{3}}\langle \langle TTT \rangle \rangle-\frac{\varepsilon_{4}\cdot k_{3}\varepsilon_{3}\cdot\varepsilon_{4}}{2k_{3}^{2}}\varepsilon_{3}^{(i}k_{3}^{j)}\partial_{\varepsilon_{3}^{ij}}\left(\left.\langle \langle TTT \rangle \rangle\right|_{k_{3}=0}-\langle \langle TTT \rangle \rangle\right)+...
\label{4ptgrsoft}
\end{equation}
In obtaining this formula we used the following identity:
\begin{equation}
\int\frac{dz}{z^{2}}\left(1+k_{1}z\right)\left(1+k_{2}z\right)e^{-k_{12}z}\left(1-\left(1+k_{3}z\right)e^{-k_{3}z}\right)=\frac{k_{3}^{2}}{k_{12}}\left(\frac{k_{1}k_{2}}{k_{123}^{2}}+\frac{k_{12}}{k_{123}}\right),
\end{equation}
and replaced $\tilde{V}_{ab}^{ij}$ with the three-graviton vertex. While the three-graviton vertex is not equal to the square
of the three-gluon vertices unless all three legs are external,
the difference between them actually doesn't contribute to the amplitude
in the soft limit, as we will prove in the next section.  

The first term in parenthesis in \eqref{4ptgrsoft} corresponds to a boundary contact term. To see this, let us write it out explicitly:
\begin{equation}
\frac{2\varepsilon_{4}\cdot k_{3}\varepsilon_{3}\cdot\varepsilon_{4}}{k_{3}^{2}}\varepsilon_{3}^{(i}k_{3}^{j)}\partial_{\varepsilon_{3}^{ij}}\left.\langle \langle TTT \rangle \rangle\right|_{k_{3}=0}=\frac{2\varepsilon_{4}\cdot k_{3}\varepsilon_{3}\cdot\varepsilon_{4}}{k_{3}^{2}}\varepsilon_{3}^{(i}k_{3}^{j)}\tilde{V}_{12}^{ij}\int_{z_{0}}^{\infty}\frac{dz}{z^{2}}\left(1+k_{1}z\right)\left(1+k_{2}z\right)e^{-k_{12}z},
\label{amputatedgr}
\end{equation}  
where we have put a cutoff on the lower limit of the $z$ integral to regulate the divergence. In more detail, this integral is given by
\begin{equation}
\int_{z_0}^{\infty}\frac{dz}{z^{2}}\left(1+k_{1}z\right)\left(1+k_{2}z\right)e^{-k_{12}z}=\frac{1}{z_0}+\frac{k_{1}k_{2}}{k_{12}}-k_{1}-k_{2}.
\label{amputatedint}
\end{equation}
Note that this divergence cancels a divergence in the second term in parenthesis in \eqref{4ptgrsoft}, which we left out of \eqref{2pt3ptgr} for simplicity. We then find that
\begin{eqnarray}
\frac{\varepsilon_{4}\cdot k_{3}\varepsilon_{3}\cdot\varepsilon_{4}}{2k_{3}^{2}}\varepsilon_{3}^{(i}k_{3}^{j)}\partial_{\varepsilon_{3}^{ij}}\left.\langle \langle TTT \rangle \rangle\right|_{k_{3}=0}=-\frac{\varepsilon_{4}\cdot k_{3}\varepsilon_{1}\cdot\varepsilon_{2}\varepsilon_{3}\cdot\varepsilon_{4}}{2k_{3}^{2}}\left(\varepsilon_{1}\cdot\varepsilon_{2}\varepsilon_{3}\cdot k_{1}+{\rm cyclic}\right)\nonumber
\\
\times \left(k_{1}^{2}-k_{2}^{2}\right)\left(\frac{1}{z_0}+\frac{k_{1}k_{2}}{k_{12}}-k_{1}-k_{2}\right).
\label{amputatedgrt}
\end{eqnarray}   
Noting that
\begin{equation}
\frac{\left(k_{1}^{2}-k_{2}^{2}\right)k_{1}k_{2}}{k_{12}}=k_{1}^{2}k_{2}-k_{2}^{2}k_{1},
\end{equation}
we see that all the terms on the right hand side of \eqref{amputatedgrt} are analytic in two momenta and therefore correspond to boundary contact terms in position space. Hence, we discard the first term in parenthesis in \eqref{4ptgrsoft} and are left with
\begin{align}
\lim_{\vec{k}_{4}\rightarrow0}\langle \langle TTTT \rangle \rangle^{(s)}=&-\frac{\left(\varepsilon_{4}\cdot k_{3}\right)^{2}}{2k_{3}}\partial_{k_{3}}\langle \langle TTT \rangle \rangle+\frac{\varepsilon_{4}\cdot\varepsilon_{3}\varepsilon_{4}\cdot k_{3}}{2k_{3}^{2}}\varepsilon_{3}^{(i}k_{3}^{j)}\partial_{\varepsilon_{3}^{ij}}\langle \langle TTT \rangle \rangle+... \label{gr4ptsoft1}\\
=&-\frac{1}{2}\varepsilon_{4}^{ij}k_{3i}\partial_{k_{3j}}\langle\langle TTT\rangle\rangle+...,
\label{gr4ptsoft}
\end{align}
where the momentum derivative in the second line acts on the bulk-to-boundary propagator of leg 3. To go from the second line to the first line, we recalled that the bulk-to-boundary propagator can be expressed in terms of a scalar propagator dressed with a polarisation and combined \eqref{momdervident} with \eqref{gravpol} to obtain \cite{Pimentel:2013gza}
\begin{equation}
\frac{\partial\varepsilon_{ij}}{\partial k^{l}}=-\frac{k_{i}\varepsilon_{jl}+k_{j}\varepsilon_{il}}{k^{2}}.
\label{grdervident}
\end{equation}
Note the similarity to the soft gluon limit in \eqref{4ptsoft}. In particular, \eqref{gr4ptsoft1} consists of an energy derivative which gives rise to a cubic energy pole which can be thought of as the AdS analogue of the Weinberg soft pole, plus a polarisation derivative which is subleading in the flat space limit. The full soft limit is obtained by summing over the s,t, and u channels. 

\section{General Multiplicity}\label{sec:generalMult}

In the previous section, we computed soft limits of four-point correlators in AdS$_4$. In this section, we will generalise these formulae to arbitrary multiplicity by computing the soft limit of general class I diagrams, which take the following form:
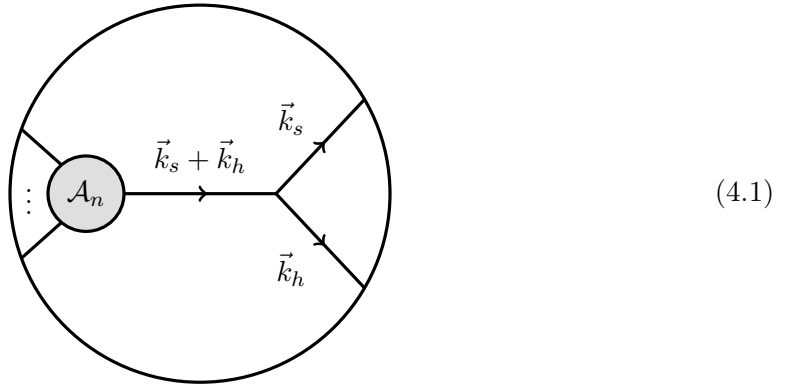
\begin{eqn}
\begin{tikzpicture}[baseline]
\draw[very thick, fill = lightgray!50] (-1.5, 0) circle (0.5);
\draw[very thick, fermion, black] (-1, 0) -- (1,0);
\draw[very thick, fermion, black] (1, 0) -- ({2.5*cos(30)}, {2.5*sin(30)});
\draw[very thick, fermion, black] (1, 0) -- ({2.5*cos(-30)}, {2.5*sin(-30)});
\draw[very thick] (0, 0) circle (2.5);

\draw[very thick, black] (-1.82, 0.38) -- ({2.5*cos(160)}, {2.5*sin(160)});
\draw[very thick, black] (-1.82, -0.38) -- ({2.5*cos(200)}, {2.5*sin(200)});
\node at (-2.25, 0) {$\vdots$};

\node at (0, 0.5) {$\vec k_s + \vec k_h$};
\node at (-1.5, 0) {$\mathcal{A}_{n}$};
\node at (1.2, 1) {$\vec k_s$};
\node at (1.2, -1) {$\vec k_h$};
\end{tikzpicture} 
\label{class1diaga}
	\end{eqn}
where $\vec{k}_s$ is the momentum that we take soft and $\vec{k}_h$ is a generic hard momentum. Note that these diagrams are the same ones that give rise to the Weinberg soft theorems in flat space. In Appendix \ref{softmomentumappendix} we compute the soft limit of general diagrams in Yang-Mills using momentum space Feynman rules and show that only class I diagrams give rise to energy derivatives or poles in the energy of individual hard legs. For gravity it is much more challenging to directly evaluate class I diagrams due to the complexity of the Feynman rules so we resort to bootstrap techniques in Mellin momentum space recently developed in \cite{Mei:2023jkb,Mei:2024abu}. In particular, the blob in \eqref{class1diaga} represents an $n$-point Mellin-momentum amplitude which we will review in the next subsection. 

\subsection{Mellin-momentum amplitudes}

For many applications, it is useful to formally represent boundary correlators in terms of a certain differential operator acting on a scalar contact diagram:
\begin{eqnarray}
	\left\langle \left\langle \mathcal{O}\left(\vec{k}_{1}\right)\ldots\mathcal{O}\left(\vec{k}_{n}\right)\right\rangle \right\rangle = \int \frac{d z}{z^{d+1}} \mathcal{A}_n(z,\vec{k}_a,\vec{\varepsilon}_a)\prod_{a=1}^n\phi_{\Delta}(k_a, z), \label{ev1e}
\end{eqnarray}
where the left-hand-side represents a generic scalar or spinning correlator. This is known as the differential representation \cite{Eberhardt:2020ewh, Gomez:2021qfd, Herderschee:2022ntr}. Spinning correlators can be represented this way because bulk-to-boundary propagators can be expressed in terms of scalar propagators dressed with polarisations \cite{Mei:2023jkb}. The differential operator $\mathcal{A}_n$ contains tensor structures constructed from external polarisations and momenta, interaction vertices dressed with additional $z$ factors, and bulk-to-bulk propagators which are formally encoded by the inverse of the differential operator in \eqref{scalareom}. In more detail, noting that a bulk-to-bulk scalar propagator satisfies  
\begin{eqnarray}
		\mathcal{D}_k^\Delta (z) G_\Delta (k,z, y)  =z^{d+1} \delta(z-y),\label{twopoint}
	\end{eqnarray}
where $\Delta$ is the scaling dimension, $(z,y)$ are two radial coordinates, and $k$ is the energy flowing through the propagator, we see that the insertion of a bulk-to-bulk propagator can be represented by acting with $(\mathcal{D}_k^\Delta (z))^{-1}$ as follows:
	\begin{eqnarray}
		\begin{aligned}
			(\mathcal{D}_k^\Delta (z))^{-1} F(z) =\int \frac{d y}{y^{d+1}} G_\Delta (k,z, y) F(y),\label{qfjkqe}
		\end{aligned}
	\end{eqnarray}
where $F(z)$ is any function. For example, the s-channel exchange of a scalar field of conformal dimension $\Delta$ and momentum $k_S\equiv|\vec{k}_1+\vec{k}_2|$ can be represented as
\begin{equation}
\int\frac{dz}{z^{d+1}}\int\frac{dx}{x^{d+1}}\phi_{\Delta_{1}}\left(k_{1},z\right)\phi_{\Delta_{2}}\left(k_{2},z\right)G_{\Delta}(k_S, z,x)\phi_{\Delta_{3}}\left(k_{3},x\right)\phi_{\Delta_{4}}\left(k_{4},x\right),
\label{scalarexample}
\end{equation}
which corresponds to having $\mathcal{A}_4(z,k_1,k_2,k_3,k_4)=1/\cD_{k_S}^\Delta$ in \eqref{ev1e}. Similarly, spinning bulk-to-bulk propagators can be expressed in terms of scalar bulk-to-bulk propagators dressed with tensor structures plus additional terms that we will not need to specify because they will automatically be captured by the bootstrap procedure that we will describe in the next subsection.
	
It is convenient to perform a Mellin transformation of the bulk-to-boundary propagators  \cite{Sleight:2019hfp,Sleight:2019mgd,Sleight:2021iix}:
\begin{eqnarray}	
\phi_{\Delta}(k, z)=\int_{-i \infty}^{+i \infty} \frac{d s}{2 \pi i} z^{-2 s+d / 2} \phi_{\Delta}(s, k),
\label{Btb2}
\end{eqnarray}
where 
\begin{equation}
\phi_{\Delta}(s, k)=\frac{\Gamma\left(s+\frac{1}{2}\left(\frac{d}{2}-\Delta\right)\right) \Gamma\left(s-\frac{1}{2}\left(\frac{d}{2}-\Delta\right)\right)}{2 \Gamma\left(\Delta-\frac{d}{2}+1\right)}\left(\frac{k}{2}\right)^{-2 s+\Delta-\frac{d}{2}}
\end{equation}
is the Mellin representation of a bulk-to-boundary propagator which satisfies 
\begin{eqnarray}
	\left(z^2 k^2+(d / 2-\Delta)^2-4 s^2\right) \phi_{\Delta}(s, k)=0.
\label{onshell}
\end{eqnarray}
 After performing Mellin transformations of the bulk-to-boundary propagators in \eqref{ev1e}, we can replace certain $z$-derivatives in $\mathcal{A}_n$ with Mellin variables and the resulting object will be referred to as a Mellin-momentum amplitude\footnote{We cannot generally replace $z$ derivatives appearing in $\mathcal{D}_k^\Delta (z)^{-1}$ with Mellin variables except in the limit $k \rightarrow 0$. When converting to Mellin variables one should also note that $f(s) k^2 \rightarrow f(s+1) \left(\frac{(d/2-\Delta)^2-4s^2}{z^2}\right)$, where $f(s)$ is any rational function of the Mellin variable $s$.}. In Mellin-momentum space, the soft limit then correponds to taking $\vec{k}_a \rightarrow 0$ along with $s \rightarrow -\frac{1}{2}\left(\frac{d}{2}-\Delta\right)$, as we explain in Appendix \ref{softmellin}.

\subsection{Bootstrap}\label{bootstrapformulasim}

We will now review some bootstrap techniques developed in \cite{Mei:2023jkb,Mei:2024abu} that will be relevant to this paper. To carry out the bootstrap, we only need to specify the three-point gluon and graviton Mellin-momentum amplitudes, which are given by
\begin{align}
		\mathcal{A}^{\mathrm{YM}}_3=&z(\varepsilon_1 \cdot \varepsilon_2 \varepsilon_3 \cdot k_1+\varepsilon_2 \cdot \varepsilon_3 \varepsilon_1 \cdot k_2 + \varepsilon_3 \cdot \varepsilon_1 \varepsilon_2 \cdot k_3),\\
		\mathcal{A}^{\mathrm{GR}}_3=&	(\mathcal{A}_3^{\mathrm{YM}})^2=z^2(\varepsilon_1 \cdot \varepsilon_2 \varepsilon_3 \cdot k_1+\varepsilon_2 \cdot \varepsilon_3 \varepsilon_1 \cdot k_2 + \varepsilon_3 \cdot \varepsilon_1 \varepsilon_2 \cdot k_3)^2.
\label{3ptmellin}
\end{align}
Higher-point amplitudes can then be bootstrapped by making an ansatz and imposing various constraints. For a general class I diagram in \eqref{class1diaga}, we make the following ansatz in Yang-Mills and gravity, respectively:
\begin{eqnarray}
\mathcal{A}_{n+1}^{\mathrm{YM}1}=\frac{a_{s+h}^{\mathrm{YM}}}{\mathcal{D}_{k_{s+h}}^{d-1}}+\frac{b_{s+h}}{k_{s+h}^{2}},\,\,\,\mathcal{A}_{n+1}^{\mathrm{GR}1}=\frac{a_{s+h}^{\mathrm{GR}}}{\mathcal{D}_{k_{s+h}}^{d}}+\frac{b_{s+h}^{(1)}}{k_{s+h}^{2}}+\frac{b_{s+h}^{(2)}}{k_{s+h}^{4}}+c^{AZ},
\label{ansatz}
\end{eqnarray}
where $k_{s+h}=|\vec{k}_s+\vec{k}_h|$ and the coefficients $a,b,c$ depend on momenta and polarisations. Note that the supercripts $\mathrm{YM}1$ and $\mathrm{GR}1$ indicate that we are only considering the Mellin-momentum representation of an individual class I diagram. The soft limit of full Mellin-momentum amplitudes will then be obtained by summing over all contributing class I diagrams. The first term in each ansatz encodes the bulk-to-bulk propagator in the class I diagram, noting that spinning bulk-to-bulk propagators can be expressed in terms of scalar bulk-to-bulk propagators dressed with tensor structures plus additional terms which we do not need to specify for the bootstrap procedure. 

The coefficients in \eqref{ansatz} can be fixed by imposing the following constraints:

\textbullet\ \textbf{Factorization}: We can choose the coefficient of $\frac{1}{\mathcal{D}^{\Delta}_{k_I}}$ to be a product of lower-point amplitudes \cite{Goodhew:2020hob,Goodhew:2021oqg,Meltzer:2020qbr}:
\begin{equation}
a_{s+h}^{\mathrm{YM}}=\sum_{r}\mathcal{A}_{3}^{{\mathrm{YM}},i}\varepsilon_{i}^{r}\varepsilon_{j}^{*r}\mathcal{A}_{n}^{{\mathrm{YM}},j},\,\,\,a_{s+h}^{\mathrm{GR}}=\sum_{r}\mathcal{A}_{3}^{{\mathrm{GR}},i_{1}i_{2}}\varepsilon_{i_{1}i_{2}}^{r}\varepsilon_{j_{1}j_{2}}^{*r}\mathcal{A}_{n}^{{\mathrm{GR}},j_{1}j_{2}},
\label{acoeffs}
\end{equation}
where $r$ labels transverse traceless states and $\mathcal{A}_n= \mathcal{A}_n^{i}\varepsilon_i$. The polarisation sums are given by
\begin{eqnarray}
\begin{aligned}
			\sum_{r} \varepsilon_i(k, r) \varepsilon_j(k, r)^* & =\eta_{ij}-\frac{k_i k_j}{k^2} \equiv \Pi_{ij} \\
			\sum_{r} \varepsilon_{ij}(k, r) \varepsilon_{kl}(k, r)^* & =\frac{1}{2} \Pi_{i k} \Pi_{j l}+\frac{1}{2} \Pi_{i l} \Pi_{k j}-\frac{1}{d-1} \Pi_{ij} \Pi_{kl}\label{projections}.
\end{aligned}
\end{eqnarray}
Note that we can add terms to the right-hand-side of \eqref{acoeffs} which are proportional to $\mathcal{D}^{\Delta}_{k_I}$, but this will cancel the $\frac{1}{\mathcal{D}^{\Delta}_{k_I}}$ in \eqref{ansatz} so such terms can be absorbed into the other coefficients of the ansatz.  
	
\textbullet\ \textbf{OPE limit}: If two operators $\mathcal{O}_a$ and $\mathcal{O}_b$ become close together in position space, in momentum space this corresponds to the sum of their momenta $\vec{k}_I=\vec{k}_a+\vec{k}_b$ becoming soft. On the other hand, the operator product expansion (OPE) implies that the correlator should factorise into a product of lower-point correlators times a two-point function which scales like $k_I^{2\Delta-d}$ (where $\Delta$ is the scaling dimension of the exchanged operator) plus terms which have no poles in $k_I$ \cite{Seery:2008ax,Leblond:2010yq,Kehagias:2012pd,Kehagias:2012td,Arkani-Hamed:2015bza,Creminelli:2012ed,Sleight:2019hfp,Bzowski:2022rlz,Assassi:2012zq,Mei:2023jkb,Mei:2024abu}. For exchanged gluons and gravitons, $\Delta=d-1$ and $d$, respectively, so for $d>2$ this contribution vanishes as $k_I \rightarrow 0$. For class I diagrams, we therefore find that
\eqs{
		\underset{k_{s+h}^2 \to 0}{\mathrm{Res}} \mathcal{A}^{\mathrm{YM}}_{n+1} = 0. \qquad \underset{k_{s+h}^{2}\to0}{\mathrm{Res}}\mathcal{A}_{n+1}^{\mathrm{GR}}=\underset{k_{s+h}^{4}\to0}{\mathrm{Res}}\mathcal{A}_{n+1}^{\mathrm{GR}}=0.
\label{opeconstraints}
}
These constraints fix the $b$'s in \eqref{ansatz}.

\textbullet\ \textbf{Adler zero}: Suppose we reduce two external gravitons to scalars by imposing $\varepsilon_a \cdot \varepsilon_b=1$ and $\varepsilon_a \cdot k_c=\varepsilon_b \cdot k_c=0$, where $a,b$ label the scalars and $c$ labels any leg. After doing so, we obtain a correlator of the form $\langle \phi \phi hhh\dots \rangle $ where the scalars couple to gravity via $\nabla^\mu \phi \nabla^\nu \phi h_{\mu \nu} $ and enjoy a shift symmetry $\phi \to \phi + \mathrm{constant}$. It follows that the Mellin-momentum amplitude must vanish when taking the soft limit of a scalar field, which is an Alder zero in curved space \cite{Armstrong:2022vgl,Armstrong:2022csc}. This constraint fixes $c^{AZ}$ in \eqref{ansatz}.

\subsection{Gluon Soft Limit} \label{gluonsoftbootstrap}

Let us now use the bootstrap method described above to compute the soft limit of a general class I diagram for gluons:	
	\begin{eqn}
	\lim_{\vec k_s \to 0}
\begin{tikzpicture}[baseline]
\draw[very thick, fill = lightgray!50] (-1.5, 0) circle (0.5);
\draw[very thick, fermion, black] (-1, 0) -- (1,0);
\draw[very thick, fermion, black] (1, 0) -- ({2.5*cos(30)}, {2.5*sin(30)});
\draw[very thick, fermion, black] (1, 0) -- ({2.5*cos(-30)}, {2.5*sin(-30)});
\draw[very thick] (0, 0) circle (2.5);


\draw[very thick, black] (-1.82, 0.38) -- ({2.5*cos(160)}, {2.5*sin(160)});
\draw[very thick, black] (-1.82, -0.38) -- ({2.5*cos(200)}, {2.5*sin(200)});

\node at (-2.25, 0) {$\vdots$};

\node at (0, 0.5) {$\vec k_s + \vec k_h$};

\node at (-1.5, 0) {$\mathcal{A}^{\mathrm{YM}}_{n}$};
\node at (1.2, 1) {$\vec k_s$};
\node at (1.2, -1) {$\vec k_h$};
\end{tikzpicture} 
\label{class1diagym}
	\end{eqn}
Starting from the ansatz in \eqref{ansatz}, we have
\begin{eqnarray}
		\mathcal{A}^{\mathrm{YM1}}_{n+1}=\frac{a^{\mathrm{YM}}_{s+h}}{\mathcal{D}^{d-1}_{k_{s+h}}}+\frac{b_{s+h}^\mathrm{YM}}{k_{s+h}^2},\label{ymansatz}
\end{eqnarray}
where $k_{s+h}=|\vec{k}_s+\vec{k}_h|$. Factorisation then implies that
	\begin{eqnarray}
		a_{s+h}^{\mathrm{YM}}=\sum_{r}\mathcal{A}_{3}^{{\mathrm{YM}},i}\varepsilon_{i}^{r}\varepsilon_{j}^{r,*}\mathcal{A}_{n}^{{\mathrm{YM}},j}=\mathcal{A}_{3}^{{\mathrm{YM}},i}\Pi_{ij}\mathcal{A}_{n}^{{\mathrm{YM}},j},\label{firstBT}
	\end{eqnarray}
where the arguments of the three-point subamplitude are given by $\mathcal{A}_3^{\mathrm{YM}}(\varepsilon_h,\varepsilon_s,\varepsilon_r,k_h,k_s,-k_{s+h}) $ and $\Pi_{ij}$ is given in \eqref{projections}. If we now take $\vec{k}_s$ soft, we obtain
	\begin{eqnarray}
		\lim_{\vec{k}_{s}\to0}\frac{a_{s+h}}{\mathcal{D}_{k_{s+h}}^{d-1}}=-\frac{z \varepsilon_{s}\cdot k_{h}\mathcal{A}_{n}^{{\mathrm{YM}}}}{\mathcal{D}_{k_{h}}^{d-1}},\label{BTsoftlimit}
	\end{eqnarray}
where we used the explicit form of the three-gluon amplitude in \eqref{3ptmellin}.

Next let us use the OPE constraint in \eqref{opeconstraints} to fix $b_{s+h}$ in \eqref{ymansatz}. Plugging \eqref{firstBT} into \eqref{ymansatz} and noting that $\Pi_{ij}$ has a pole in $k_{s+h}^2$, we find that
	\begin{eqnarray}
		b_{s+h}\eqnmy -\underset{k_{s+h}^{2}\to0}{\mathrm{Res}}\frac{\mathcal{A}_{3}^{{\mathrm{YM}},i}\Pi_{ij}\mathcal{A}_{n}^{{\mathrm{YM}},j}}{\mathcal{D}_{k_{s+h}}^{d-1}} \nonmy
		\eqnmy   \frac{z ({k}_{s}^{j}+{k}_{h}^{j})\mathcal{A}_{n}^{{\mathrm{YM}},j}(k_{h}^{2}-k_{s}^{2})\varepsilon_{s}\cdot\varepsilon_{h}}{4\left(d-2s_{h}-2s_{s}\right)\left(s_{h}+s_{s}-1\right)}\nonmy
		\eqnmy   \frac{{k}_{h+s}^{j}\partial_{\varepsilon_{h}^{j}}\mathcal{A}_{n}^{\mathrm{YM}}\varepsilon_{s}\cdot\varepsilon_{h}(s_{h}-s_{s})}{2 z\left(\frac{d}{2}-\left(1+s_{h}+s_{s}\right)\right)}.
\label{bsh} 
	\end{eqnarray}
To obtain the second line in \eqref{bsh}, we used 
\begin{eqnarray}\label{OpDmellin}
			\lim_{\vec{k}_{s+h}\to0}\frac{1}{\mathcal{D}_{k_{s+h}}^{d-1}}=\frac{1}{2\left(d-2 s_h-2 s_s\right)\left(s_h+s_s-1\right)},
\end{eqnarray}
which follows from acting with $\mathcal{D}^{-1}$ on the Mellin representation \eqref{Btb2} of $z\phi_{d-1}\left(z, k_h\right) \phi_{d-1}\left(z, k_s\right)$. To obtain the third line in \eqref{bsh}, we used $f(s) k^2 \rightarrow f(s+1)\left(\frac{(d / 2-\Delta)^2-4 s^2}{z^2}\right)$. Using \eqref{bsh} and recalling that we must take $s \rightarrow \frac{d}{4}-1/2$ in the soft limit, we find that
	\begin{eqnarray}
		\lim_{\vec{k}_{s}\to0}\frac{b_{s+h}}{k_{h+s}^{2}}=-\varepsilon_{s}\cdot\varepsilon_{h}\frac{k_{h}\cdot\partial_{\varepsilon_{h}}\mathcal{A}_{n}^{{\mathrm{YM}}}}{2 z k_{h}^{2}}.
\label{bsoft}
	\end{eqnarray}
Combining the results in \eqref{BTsoftlimit} and \eqref{bsoft}, we find that the soft limit of a general class I diagram in YM is given by
\begin{equation}
\lim_{\vec{k}_{s}
	\rightarrow
	0}
	\mathcal{A}_{n+1}^{\mathrm{YM1}}=-\frac{z \varepsilon_{s}\cdot k_{h}\mathcal{A}_{n}^{\mathrm{YM}}}{\mathcal{D}_{k_{h}}^{d-1}}-\varepsilon_{s}\cdot\varepsilon_{h}\frac{ k_h \cdot \partial_{\varepsilon_h}\mathcal{A}_{n}^{\mathrm{YM}}}{2 z k_{h}^{2}}.	
\label{ymres}
\end{equation}
The first term is the natural generalization of a Weinberg soft term in flat space after replacing $ k_s\cdot k_{h} \to \mathcal{D}_{k_{s+h}}^{\Delta} $, while the second term is subleading in the flat space limit which can be seen by taking $z \rightarrow \infty$. 

In Appendix \ref{meltomo} we show how to translate \eqref{ymres} to correlation functions in momentum space. After relabeling the soft leg as $n+1$ and summing over all class I diagrams, we then obtain the following formula for the soft limit of color-ordered gluon correlators in AdS$_{d+1}$:
\begin{align}
\lim _{\vec{k}_{n+1} \rightarrow 0} \langle  \langle J \ldots J\rangle\rangle_{n+1}&=
\mathcal{N}_{d-1} \left\{ \frac{\varepsilon_{n+1} \cdot k_n}{2 k_n} \partial_{k_n} \langle  \langle J \ldots J\rangle\rangle_n-\frac{\varepsilon_{n+1} \cdot \varepsilon_n}{2 k_n^2} k_n \cdot \partial_{\varepsilon_n} \left\langle\langle J\ldots J\rangle\right\rangle_n\right\} \nonumber \\	
-&\left(\vec{k}_{n}\rightarrow\vec{k}_{1},\vec{\varepsilon}_{n}\rightarrow\vec{\varepsilon}_{1}\right)+... \label{ymsoftabc}\\
&=\frac{\mathcal{N}_{d-1}}{2}
\left\{ \varepsilon^i_{n+1}  \partial_{k_{ni}}\langle\langle J \ldots J\rangle\rangle_n- \varepsilon^i_{n+1} \partial_{k_{1i }}\langle\langle J \ldots J\rangle\rangle_n\right\}+\ldots
\label{wqkjhf}
\end{align}
where $\mathcal{N}_{d-1}=\frac{2^{(d-3)/2}\Gamma\left(\frac{d-2}{2}\right)}{\sqrt{\pi}}$ and ... denotes contributions from non-class I diagrams. Note that the derivatives only act on the bulk-to-boundary propagators of hard legs. In particular, the energy derivative in \eqref{ymsoftabc} gives rise to a double pole in the energy while the polarisation derivative is subleading in the flat space limit. These two terms can then be combined into a single momentum derivative acting on bulk-to-boundary propagators using \eqref{momdervident}. In Appendix \ref{softmomentumappendix}, we compute the soft limit of general YM diagrams using momentum space Feynman rules and find that non-class I diagrams do not give energy derivatives or poles in the energy of individual hard legs. Hence, \eqref{ymsoftabc} provides non-trivial constraints on gluon correlators. We illustrate how this works at five points in Appendix \ref{5ptchecks} and the attached Mathematica notebook \texttt{5ptYM.nb}. 

Note that the soft factors of  \eqref{wqkjhf} are infrared finite, in contrast to the Weinberg soft factors in flat space. Another important property of Weinberg soft factors is that they are gauge-invariant \cite{Weinberg:1965nx}. On the other hand, the soft factors in \eqref{wqkjhf} do not exhibit this property because under gauge transformations correlators get shifted by boundary contact terms (see for example \cite{Maldacena:2011nz,Baumann:2020dch,Armstrong:2020woi} for more details). Moroever, \eqref{wqkjhf} only encodes the contribution of class I diagrams to the soft limit. In the next section we will see that similar comments apply to graviton correlators. 

\subsection{Graviton Soft Limit}

As we did for gluons, we will now consider the soft limit of class I diagrams for gravitons, which we depict below:
\begin{eqn}
	\lim_{\vec k_s \to 0}
\begin{tikzpicture}[baseline]
\draw[very thick, fill = lightgray!50] (-1.5, 0) circle (0.5);
\draw[very thick, fermion, black] (-1, 0) -- (1,0);
\draw[very thick, fermion, black] (1, 0) -- ({2.5*cos(30)}, {2.5*sin(30)});
\draw[very thick, fermion, black] (1, 0) -- ({2.5*cos(-30)}, {2.5*sin(-30)});
\draw[very thick] (0, 0) circle (2.5);


\draw[very thick, black] (-1.82, 0.38) -- ({2.5*cos(160)}, {2.5*sin(160)});
\draw[very thick, black] (-1.82, -0.38) -- ({2.5*cos(200)}, {2.5*sin(200)});

\node at (-2.25, 0) {$\vdots$};

\node at (0, 0.5) {$\vec k_s + \vec k_h$};

\node at (-1.5, 0) {\scriptsize$\mathcal{A}^{\text{GR}}_{n}$};
\node at (1.2, 1) {$ \vec{k}_s$};
\node at (1.2, -1) {$ \vec{k}_h$};
\label{class1diaggr}
\end{tikzpicture} 
\end{eqn}
In this case, we start with the ansatz 
\begin{eqnarray}
\mathcal{A}_{n+1}^{\mathrm{GR1}}=	\frac{a^{\mathrm{GR}}_{s+h}}{\mathcal{D}_{k_{s+h}}^{d}}+\frac{b^{(1)}_{s+h}}{k_{s+h}^2}+\frac{b^{(2)}_{s+h}}{k_{s+h}^4}+c^{AZ},
\label{gransatz}
\end{eqnarray}
where the first term is determined from factorization, the second and the third terms from the OPE limit, and the fourth term from the Adler zero condition after dimensional reduction. We have included a Mathematica file \texttt{Class1.nb} with more details, so we will just sketch how this works below. As explained in \eqref{acoeffs}, factorisation implies that 
\begin{equation}
a_{s+h}^{\mathrm{GR}}=\sum_{r}\mathcal{A}_{3}^{{\mathrm{GR}},i_{1}i_{2}}\left(\frac{1}{2}\Pi_{i_{1}j_{1}}\Pi_{i_{2}j_{2}}+\frac{1}{2}\Pi_{i_{1}j_{2}}\Pi_{j_{1}i_{2}}-\frac{1}{d-1}\Pi_{i_{1}i_{2}}\Pi_{j_{1}j_{2}}\right)\mathcal{A}_{n}^{{\mathrm{GR}},j_{1}j_{2}},
\label{grbootstrapa1}
\end{equation}
where the sum is over transverse traceless states. Taking the soft limit $\vec{k}_s\to 0$ then gives
\begin{eqnarray}
\lim_{\vec{k}_{s}\rightarrow0}\frac{a_{s+h}^{\mathrm{GR}}}{\mathcal{D}_{k_{s+h}}^{d}}=\frac{z^{2}\left(\varepsilon_{s}\cdot k_{h}\right)^{2}\mathcal{A}_{n}^{{\mathrm{GR}}}}{\mathcal{D}_{k_{h}}^{d}}.
\label{ares}
\end{eqnarray}

Next let us use the OPE constraints in \eqref{opeconstraints} to fix $b^{(1)}_{s+h}$ and $b^{(2)}_{s+h}$ in \eqref{gransatz}. Since $\Pi_{ij}$ has a pole in $k_{s+h}^2$, we see that the right-hand-side of \eqref{grbootstrapa1} has both $k_{s+h}^2$ and $k_{s+h}^4$ poles. Hence, the OPE constraints imply that   
\begin{eqnarray}
		\frac{b_{s+h}^{(1)}}{k_{s+h}^2}\eqnmy-\frac{1}{ k_{h+s}^2}\left(\underset{k_{s+h}^2\to 0}{\mathrm{Res}}\ 	\frac{a_{s+h}}{\mathcal{D}_{k_{s+h}}^d}\right)\nonmy  \frac{b_{s+h}^{(2)}}{k_{s+h}^4}\eqnmy -\frac{1}{ k_{h+s}^4}\left(\underset{k_{s+h}^4\to 0}{\mathrm{Res}}	\  \frac{a_{s+h}}{\mathcal{D}_{k_{s+h}}^d}\right).
\label{opeconstraintgr}
\end{eqnarray}
Since there are $1/k_{s+h}^4$ poles, we must expand $1/\cD$ to order $k_{s+h}^{2}$ in order to compute the residue of $1/k_{s+h}^2$\footnote{One can do so using $(A+B)^{-1}=A^{-1}-A^{-1} B(A+B)^{-1}$, where $A$ and $B$ are operators.}:
\begin{eqnarray}
\frac{1}{\mathcal{D}_{k_{s+h}}^{d-1}}\eqnmy\frac{1}{2\left(s_h+s_s-1\right)\left(d-2 s_h-2 s_s+2\right)}-\nonmy&&\frac{z^2 k_{s+h}^2}{4\left(s_h+s_s-2\right)\left(s_h+s_s-1\right)\left(d-2 s_h-2 s_s+2\right)\left(d-2 s_h-2 s_s+4\right)}+...\nonmy
\label{gr1dexp}
	\end{eqnarray} 
where ... denote higher order terms in $k_{s+h}^2$ and we computed the action of $1/\cD$ on the Mellin representation \eqref{Btb2} of $z^2\phi_{d}\left(z, k_h\right) \phi_{d}\left(z, k_s\right) $. Plugging \eqref{grbootstrapa1} and \eqref{gr1dexp} into \eqref{opeconstraintgr} then determines $b^{(1)}_{s+h}$ and $b^{(2)}_{s+h}$. In the soft limit we then find
\begin{equation}
		\lim_{\vec{k}_{s}\rightarrow0}\frac{b_{s+h}^{(1)}}{k_{s+h}^{2}}+\frac{b_{s+h}^{(2)}}{k_{s+h}^{4}}=-\frac{\left(\varepsilon_{s}\cdot\varepsilon_{h}\right)^{2}\mathcal{A}_{j,n}^{j,{\mathrm{GR}}}}{4(1-d)}+\frac{\varepsilon_{s}\cdot\varepsilon_{h}\varepsilon_{s}\cdot k_{h}\varepsilon_{h}^{i}k_{h}^{j}\mathcal{A}_{n}^{{\mathrm{GR}},ij}}{k_{h}^{2}},
\label{midres}
\end{equation}
where we recalled that we must take the Mellin variable $s \rightarrow d/4$ in this limit.

Finally, let us deduce the coefficient $c^{AZ}$ in \eqref{gransatz} by imposing $\varepsilon_s \cdot \varepsilon_h=1, \varepsilon_s \cdot k_a=0, \varepsilon_h \cdot k_a=0$, where $a$ can be any leg. This reduces the two external legs attached to the three-point vertex in \eqref{class1diaggr} to scalars which enjoy shift symmetries and therefore exhibit Adler zeros. Demanding that the soft limits of these legs vanishes then fixes $c^{AZ}$ (see \texttt{Class1.nb} for the expression). In the soft limit we then find that $c^{AZ}$ reduces to
	\begin{eqnarray}
		\lim_{\vec{k}_{s}\rightarrow0}c^{AZ}=\frac{\left(\varepsilon_s \cdot \varepsilon_h\right)^2 \mathcal{A}_{j, n}^{j,\mathrm{GR}}}{4(1-d)}.\label{adlerzero}
	\end{eqnarray}
Adding the results in \eqref{ares}, \eqref{midres}, and \eqref{adlerzero} finally gives
\begin{equation}
	\lim_{\vec{k}_{s}\rightarrow0}\mathcal{A}_{n+1}^{\mathrm{GR1}}=\frac{z^{2}\left(\varepsilon_{s}\cdot k_{h}\right)^{2}\mathcal{A}_{n}^{\mathrm{GR}}}{\mathcal{D}_{k_{h}}^{d}}+\frac{\varepsilon_{s}\cdot\varepsilon_{h}\varepsilon_{s}\cdot k_{h}\varepsilon_{h}^{i}k_{h}^{j}\mathcal{A}_{n}^{\mathrm{GR},ij}}{k_{h}^{2}}.
\label{grres}
	\end{equation}
Note the similarity to the gluonic soft limit in \eqref{ymres}: the first term contains the soft limit of a bulk-to-bulk scalar propagator and represents the analogue of a Weinberg soft term in AdS while the second term contains a momentum derivative is subleading in the flat space limit, $z\rightarrow \infty$. 

In Appendix \ref{meltomo} we explain how to translate \eqref{grres} to correlators in momentum space. Relabelling the soft leg as $n+1$ and summing over all class I diagrams then gives the following formula for the soft limit of graviton correlators in AdS$_{d+1}$:
\begin{align}
\lim _{\vec{k}_{n+1} \rightarrow 0}\langle \langle T \ldots T\rangle\rangle_{n+1} 
		= &\mathcal{N}_{d} \Big\{ \sum_{a= 1}^n -\frac{(\varepsilon_{n+1} \cdot k_a)^2}{2 k_a} \partial_{k_a}\langle\langle T \ldots T\rangle\rangle_n  \nonumber  \\
&+\frac{\varepsilon_{n+1} \cdot \varepsilon_a\varepsilon_{n+1} \cdot k_a}{2 k_a^2}\varepsilon_a^{(i} k_a^{j)}\partial_{\varepsilon_{a}^{ij}}\langle\langle T \ldots T\rangle\rangle_n\Big\}+\ldots, \label{grsoft12} \\
=&-\frac{\mathcal{N}_{d}}{2}\sum_{a=1}^{n}\varepsilon_{n+1}^{ij}k_{ai}\partial_{k_{aj}}\langle\langle T...T\rangle\rangle_{n}+...,
\label{grsoftresmom}
\end{align}
where $\mathcal{N}_{d}=\frac{2^{(d-1)/2}\Gamma\left(\frac{d}{2}\right)}{\sqrt{\pi}}$, the derivatives only act on bulk-to-boundary propagators of the hard legs, and ... denote contributions from non-class I diagrams. As in YM, those contributions do not contain energy derivatives or poles in the energy of individual hard legs. Hence, the soft limit of class I diagrams can be distinguished from that of other diagrams and \eqref{grsoft12} imposes non-trivial constraints on graviton correlators. Note that the energy and polarisation derivatives in \eqref{grsoft12} can be combined into a single momentum derivative in the third line using \eqref{momdervident} and \eqref{grdervident}.

\section{Conclusion} \label{conclusion}

In this paper we derived formulae for the soft limit of gluon and graviton correlators in AdS arising from class I diagrams which give rise to the Weinberg soft theorems in flat space. At four points, we obtained these formulae using explicit formulae recently obtained using cosmological boostrap methods and the double copy \cite{Bonifacio:2022vwa,Armstrong:2023phb}. We then generalised them to arbitrary multiplicity by computing the soft limit of general class I diagrams using bootstrap methods in Mellin-momentum space which were recently developed in \cite{Mei:2023jkb,Mei:2024abu}. The soft limits take the form of a differential operator acting on bulk-to-boundary propagators of lower-point correlators which is schematically a sum of two terms: an energy derivative dressed with the same tensor structure appearing in the Weinberg soft theorems plus a polarisation derivative which is subleading in the flat space limit. These two terms can be combined into a single momentum derivative acting on hard bulk-to-boundary propagators. Other classes of diagrams can also contribute to the soft limit above four points, but they have different pole structure than class I diagrams. In particular, they do not exhibit energy derivatives, which give rise to higher order energy poles, or poles in the energy of individual hard legs. Our soft limit formulae therefore provide useful constraints on gluon and graviton correlators. 

There are a number of follow-up directions. First of all, it would be very interesting to understand how to relate our soft limit formulae to consistency conditions for inflationary correlators, which were derived from Ward identities associated with certain large diffeomorphism symmetries \cite{Creminelli:2012ed,Hinterbichler:2012nm}. The first step would be to adapt these Ward identities from in-in correlators to wavefunction coefficients which can be mapped to AdS boundary correlators by analytic continuation. Given that the Weinberg soft theorems for scattering amplitudes can be derived from Ward identities associated with BMS symmetry \cite{Strominger:2013jfa}, it seems plausible that the soft gluon and graviton formulae in this paper can be derived from an analogue of BMS symmetry recently discovered in (A)dS known as $\Lambda$-BMS symmetry \cite{Compere:2019bua}. It would also be interesting to understand how these symmetries are related to those which arise in the flat space limit (recent work showing how soft factors and BMS symmetry arise from the flat space limit of conformal Ward identities \cite{Hijano:2020szl,deGioia:2023cbd} may be relevant for this purpose).  

It would also be interesting to extend our calculations to subleading order in the soft momentum. These could in turn be used to deduce new consistency relations on inflationary correlators. In flat space, non-class I diagrams play an important role in subleading soft theorems and multiparticle soft limits \cite{Chakrabarti:2017ltl}, so we expect them to play an important role in AdS as well. In order to investigate the universality of soft gluon and graviton limits in AdS, we should also consider couplings to various kinds of matter \cite{chandra}. Recently, soft limits of certain supersymmetric correlators in AdS$_5$ were investigated in \cite{Green:2020eyj,Cao:2024bky} so it would be interesting to make contact with those results as well. Finally, it would be of interest to see if soft limits can be used to bootstrap higher-point correlators. Indeed, soft limits impose very powerful constraints on scattering amplitudes and in some cases can even fully determine them \cite{Boucher-Veronneau:2011rwd}. For example, a certain class of gluon amplitudes known as MHV amplitudes can be fully reconstructed from their soft limits \cite{Britto:2005fq,Arkani-Hamed:2012zlh} and are described by a very concise expression for any multiplicity known as the Parke-Taylor formula \cite{Parke:1986gb}. If we can use similar reasoning derive an all-multiplicity formula for gluon correlators in AdS that would be very significant. Twistor string formulae along these lines have been proposed in \cite{Adamo:2012nn,Adamo:2015ina,Adamo:2024hme}, although their physical interpretation is not yet clear. We hope to report on these exciting directions in the future.

\begin{center}
\textbf{Acknowledgements}
\end{center}

We thank Sadra Jazayeri, Silvia Nagy, Enrico Pajer, Guilherme Pimentel, Santiago Agui Salcedo, Charlotte Sleight, and Massimo Taronna for useful discussions.  AL is supported an STFC Consolidated Grant ST/T000708/1. JM is supported by a Durham-CSC Scholarship. CC is supported by the STFC consolidated
grant (ST/X000583/1) “New Frontiers in Particle Physics, Cosmology and Gravity”.  YM is supported by a Edinburgh Global Research Scholarship.

\appendix

\section{Four-point GR soft limit} \label{4ptgr}

In this Appendix, we will derive \eqref{4ptgrsoft1}. Our starting point will equation 4.9 of \cite{Armstrong:2023phb}, which gives the s-channel contribution to the four-point tree-level graviton correlator\footnote{We have introduced an overall factor of $1/16$.}:
\begin{align}
	\langle	\langle TTTT\rangle\rangle^{(s)}= & \frac{1}{16}{\left(\varepsilon_1 \cdot \varepsilon_2\right)^2\left(\varepsilon_3 \cdot \varepsilon_4\right)^2} {\psi_{\phi,\mathrm{DC}}^{(s)}}+\frac{1}{16}\left(8\left(\varepsilon_1 \cdot \varepsilon_2\right)\left(\varepsilon_3 \cdot \varepsilon_4\right) W_s k_s^2 \Pi_{1,1}+{16 W_s^2}\right) f_{2,2} \nonumber \\
		& -\frac{1}{16}\left(\varepsilon_1 \cdot \varepsilon_2\right)\left(\varepsilon_3 \cdot \varepsilon_4\right) k_{12} k_{34}\left({8 W_s \Pi_{1,0}}+{\alpha \beta V_c^s}\right) f_{2,1} \nonumber \\
		& +\frac{1}{16}\left(\left(V_c^s\right)^2+\frac{1}{2}{\left(\varepsilon_1 \cdot \varepsilon_2\right)^2\left(\varepsilon_3 \cdot \varepsilon_4\right)^2}\right) {f_a} +\nonumber  \\
		&+\frac{1}{16}\left({\left(\varepsilon_1 \cdot \varepsilon_2\right)\left(\varepsilon_3 \cdot \varepsilon_4\right)\left(\vec{k}_1-\vec{k}_2\right) \cdot\left(\vec{k}_3-\vec{k}_4\right)}+{8 W_s}\right) V_c^s f_b.
\label{schannelGRsoft}
\end{align}
where $W_s, \Pi_{1,1}, \Pi_{1,0}, V_c^s$ are given in \eqref{definitions}, $\alpha=k_1-k_2$, $\beta=k_3-k_4$, and
\begin{align}
	\psi_{\phi, \mathrm{DC}}^{(s)}= & \frac{1}{3} k_S^4 f_{2,2} \Pi_{2,2}-\frac{1}{3} k_S^2 k_{12} k_{34} f_{2,1} \Pi_{2,1}+\frac{1}{2} f_{2,0} \frac{k_{12}^2 \alpha^2 k_{34}^2 \beta^2}{k_S^4}  \nonumber\\
& -\frac{1}{2} f_{2,1}\left(\left(k_{12}^2+\alpha^2-k_s^2-\frac{k_{12}^2 \alpha^2}{k_S^2}\right) \frac{k_{34}^2 \beta^2}{k_S^2}+\frac{k_{12}^2 \alpha^2}{k_S^2}\left(k_{34}^2+\beta^2-k_s^2-\frac{k_{34}^2 \beta^2}{k_S^2}\right)\right)
 ,\nonumber\\
f_{2,2}&={\frac{2 k_1 k_2 k_3 k_4\left(E_L E_R+k_{1234} k_S\right)}{k_{1234}^3 E_L^2 E_R^2}}+\frac{k_1 k_2\left(E_L k_{34}+k_{1234} k_S\right)}{k_{1234}^2 E_L^2  E_R}
+	\nonumber\\
&\qquad
{\frac{k_3 k_4\left(k_{1234} k_S+
		E_R k_{12}\right)}{k_{1234}^2 E_L  E_R^2}}+\frac{E_L E_R-k_S^2}{k_{1234} E_L E_R}, \nonumber \\
f_{2,1}&=\frac{2 k_1 k_3 k_4 k_2}{k_{1234}^3 k_{12} k_{34}}+\frac{k_1 k_2}{k_{1234}^2 k_{12}}+\frac{k_3 k_4}{k_{1234}^2 k_{34}}+\frac{1}{k_{1234}}, \nonumber \\
f_a &=\left(k_{12} k_{34}+k_S^2\right) f_b+\frac{1}{k_{1234}}\left(
2 k_1 k_2 k_3 k_4-k_1 k_2\left(2 k_{1234}^2+k_{12}^2\right)
\right)+ \nonumber \\
& \qquad \frac{1}{k_{1234}}\left({-k_3 k_4\left(2 k_{1234}^2+k_{34}^2\right)-2 k_{12} k_{34} k_{1234}^2+k_{1234}^4}\right) , \nonumber \\
f_b &=\left(\frac{2 k_1 k_2 k_3 k_4}{k_{1234}^3}+k_1 k_2 \frac{k_{34}+k_{1234}}{k_{1234}^2}+k_3 k_4 \frac{k_{12}+k_{1234}}{k_{1234}^2}+\frac{k_{12} k_{34}-k_{1234}^2}{k_{1234}}\right),\nonumber\\
\Pi_{2,2}&=\frac{3}{2 k_S^4}\left(\vec{k}_1-\vec{k}_2\right)^i\left(\vec{k}_1-\vec{k}_2\right)^j\left(\Pi_{i l} \Pi_{j m}+\Pi_{i m} \Pi_{j l}-\Pi_{i j} \Pi_{l m}\right)\left(\vec{k}_3-\vec{k}_4\right)^l\left(\vec{k}_3-\vec{k}_4\right)^m,\nonumber\\ 
\Pi_{2,1}&=\frac{3}{2 k_S^2 k_{12} k_{34}}\left(\vec{k}_1-\vec{k}_2\right)^i\left(\vec{k}_1-\vec{k}_2\right)^j\left(\Pi_{i l} \hat{k}_j \hat{k}_m+\Pi_{j m} \hat{k}_i \hat{k}_l+\Pi_{i m} \hat{k}_j \hat{k}_l+\Pi_{j l} \hat{k}_i \hat{k}_m\right)\nonumber\\ &\qquad \left(\vec{k}_3-\vec{k}_4\right)^l\left(\vec{k}_3-\vec{k}_4\right)^m.
\end{align}
where $k_{S}=\left|\vec{k}_{1}+\vec{k}_{2}\right|$, $E_{L}=k_{12}+k_{S}$, $E_{R}=k_{34}+k_{S}$, and $\hat{k}_i=\frac{\left(\vec{k}_1+\vec{k}_2\right)_i}{k_S}$.

Let us now take the limit $\vec{k}_4 \rightarrow 0$. We then find that $\Pi_{1,1}$ vanishes and $f_{a}$ vanishes up to boundary contact terms:
\begin{equation}
\lim_{\vec{k}_{4}\rightarrow0}f_{a}=k_{1}^{3}+k_{2}^{3}.
\end{equation}
Hence, we can drop terms proportional to $\Pi_{1,1}$ and $f_{a}$. Moreover, with a bit of algebra we can show that all terms proportional to $\left(\varepsilon_1 \cdot \varepsilon_2\right)^2\left(\varepsilon_3 \cdot \varepsilon_4\right)^2 $ vanish in the soft limit. This can understood from an emergent shift symmetry after dimenisonal reduction, as explained in section \ref{bootstrapformulasim}. We also find that the following linear combination of terms reduces to a boundary contact term in the soft limit: 
\begin{align}
	\lim_{\vec{k}_{4}\rightarrow0} & \left[-\left(\varepsilon_{1}\cdot\varepsilon_{2}\right)  \left(\varepsilon_{3}\cdot\varepsilon_{4}\right)k_{12}k_{34}\left(\alpha\beta V_{c}^{s}\right)f_{2,1}+\left(\left(\varepsilon_{1}\cdot\varepsilon_{2}\right)\left(\varepsilon_{3}\cdot\varepsilon_{4}\right)\left(\vec{k}_{1}-\vec{k}_{2}\right)\cdot\left(\vec{k}_{3}-\vec{k}_{4}\right)\right)V_{c}^{s}f_{b}\right]\nonumber \\
 &= \left(\varepsilon_1 \cdot \varepsilon_2\right)\left(\varepsilon_3 \cdot \varepsilon_4\right)V_c^s (k_1^3-k_2^3),
\end{align}
where we noted that
\begin{align}
\lim_{\vec{k}_{4}\rightarrow0}f_{2,1}&=\frac{k_{1}k_{2}}{k_{123}^{2}k_{12}}+\frac{1}{k_{123}}, \nonumber \\
	\lim_{\vec{k}_{4}\rightarrow0}f_{b}&= \frac{k_2 k_3 k_1}{k_{123}^2}+\frac{k_1 k_2+k_3 k_2+k_1 k_3}{k_{123}}-k_{123}.
\label{limits}
\end{align}
In summary, we find that the following terms in \eqref{schannelGRsoft} survive in the soft limit:
\begin{equation}
\lim_{\vec{k}_{4}\rightarrow0} \langle	\langle TTTT\rangle\rangle^{(s)} =\lim_{\vec{k}_{4}\rightarrow0}\left[W_{s}^{2}f_{2,2}-\frac{1}{2}\left(\varepsilon_{1}\cdot\varepsilon_{2}\right)\left(\varepsilon_{3}\cdot\varepsilon_{4}\right)k_{12}k_{34}W_{s}\Pi_{1,0}f_{2,1}+\frac{1}{2}W_{s}V_{c}^{s}f_{b}\right].
\end{equation}
We can now evaluate the right-hand-side using \eqref{limits} and the following soft limits:
\begin{align}
\lim_{\vec{k}_{4}\rightarrow0}W_{s}& = \left( \varepsilon_1 \cdot \varepsilon_2k_2 \cdot \varepsilon_3+\varepsilon_3 \cdot \varepsilon_1k_1 \cdot \varepsilon_2+\varepsilon_3 \cdot \varepsilon_2k_3 \cdot \varepsilon_1\right) k_3\cdot \varepsilon_4, \nonumber \\
\lim_{\vec{k}_{4}\rightarrow0}f_{2,2}&=\frac{k_{1}k_{2}}{k_{123}^{3}}+\frac{k_{1}+k_{2}}{2k_{123}^{2}}+\frac{1}{2k_{123}}.
\end{align}
We then obtain \eqref{4ptgrsoft1} after some algebra.


\section{Soft limits in Mellin space} \label{softmellin}

Using the differential representation in \eqref{ev1e}, the soft limit of a correlator can be directly evaluated by taking the soft limit of a bulk-to-boundary propagator in the contact diagram \cite{Armstrong:2022vgl}. Let us therefore consider the soft limit of a scalar bulk-to-boundary propagator in \eqref{Btb}:
\begin{eqnarray}
		\lim\limits_{k\to 0}\phi_{\Delta}(k, z)=\sqrt{\frac{2}{\pi}}\left[	2^{-\frac{d}{2}+\Delta-1} \Gamma\left(\Delta-\frac{d}{2}\right) z^{d-\Delta}+2^{\frac{d}{2}-\Delta-1} z^{\Delta} \Gamma\left(\frac{d}{2}-\Delta\right) k^{2 \Delta-d}\right].\nonmy\label{qjgew}
	\end{eqnarray}
For $\Delta>d/2$ the second term gets power suppressed leaving us with
\begin{equation}
\lim\limits _{k\to0}\phi_{\Delta}(k,z)=\sqrt{\frac{2}{\pi}}2^{-\frac{d}{2}+\Delta-1}\Gamma\left(\Delta-\frac{d}{2}\right)z^{d-\Delta}.
\end{equation}
On the other hand in Mellin space, the power expansion in \eqref{qjgew} is encoded by the residues of the integral in \eqref{Btb2}:
	\begin{eqnarray}
		\phi_{\Delta}(k, z)=\int_{-i \infty}^{+i \infty} \frac{d s}{2 \pi i} z^{-2 s+d / 2} \frac{\Gamma\left(s+\frac{1}{2}\left(\frac{d}{2}-\Delta\right)\right) \Gamma\left(s-\frac{1}{2}\left(\frac{d}{2}-\Delta\right)\right)}{2 \Gamma\left(\Delta-\frac{d}{2}+1\right)}\left(\frac{k}{2}\right)^{-2 s+\Delta-\frac{d}{2}}.\label{btbMellin}
	\end{eqnarray}
The poles are located at $s=-x+\frac{1}{2}\left(\frac{d}{2}-\Delta\right)$, corresponding to $k^{2\Delta-d+2x}$, and $s=-x-\frac{1}{2}\left(\frac{d}{2}-\Delta\right)$, corresponding to $k^{2x}$, where $x$ is a non-negative integer. For $\Delta>d/2$, the $x=0$ pole gives the leading contribution in the soft limit. Hence, in the soft limit we must take
	\begin{eqnarray}
		s \rightarrow -\frac{1}{2}\left(\frac{d}{2}-\Delta\right).
	\end{eqnarray}
In particular, for YM ($\Delta=d-1$) and GR ($\Delta=d$) we have 
\begin{eqnarray}
		\color{black}\text{YM: } s &\rightarrow& \frac{d-2}{4},\nonmy \color{black}\text{GR: } s &\rightarrow&  \frac{d}{4}.\label{softs}
	\end{eqnarray}

\section{From Mellin to Momentum Space} \label{meltomo}

In this Appendix, we will translate the soft limits for Mellin-momentum amplitudes in \eqref{ymres} and \eqref{grres} to correlators in momentum space.

\subsection{Yang-Mills}

Let us consider the first term on the right-hand-side of \eqref{ymres}. We shall denote this contribution to the soft limit by the subscript 1. Plugging this into \eqref{ev1e} to go from the Mellin momentum amplitude to the momentum space correlator gives 
\begin{eqnarray}
		\left\langle\langle J_1\ldots J_{n-1}J_hJ_s \right \rangle \rangle_1 \eqnmy -\int_0^{\infty} \frac{d z_1}{z_1^{d+1}} \frac{d z_2}{z_2^{d+1}} \phi_{\Delta}\left(z_1, k_h\right)\phi_{\Delta}\left(z_1, k_s\right) z_1 G_\Delta\left(z_1, z_2, k_h \right)  \nonmy&&\quad \times \ \varepsilon_s \cdot k_h \varepsilon_{h}^j\mathcal{A}_{n}^{{\mathrm{YM}},j}\left(z_2\right) \prod\limits_{a=1}^{n-1}\phi_{\Delta}\left(z_2, k_a\right),
\label{qiuwq}
\end{eqnarray}
where $\Delta=d-1$ and the $z_1$ in the first line comes from the definition of the three-point YM amplitude in \eqref{3ptmellin}. The scalar bulk-to-boundary propagators are given by \eqref{Btb} and the soft propagator is expanded to first order in the soft momentum:    \begin{eqnarray}
		\phi_{d-1}\left(z,k_{s}\right)=\mathcal{N}_{d-1}z+O\left(k_{s}\right),\,\,\,\mathcal{N}_{d-1}=\frac{2^{(d-3)/2}\Gamma\left(\frac{d-2}{2}\right)}{\sqrt{\pi}}.
\label{btBs1}
	\end{eqnarray}
For $d=2$ there is a subtlety because $\Gamma\left(\frac{d-2}{2}\right)$ is singular, so we will focus on $d>2$. The scalar bulk-to-bulk propagator in \eqref{qiuwq} is given by 
\begin{eqnarray}
	G_{\Delta}\left(z_{1},z_{2},k_{h}\right)=\int_{0}^{\infty}dw\frac{wz_{1}^{d/2}z_{2}^{d/2}J_{\Delta-d/2}\left(wz_{1}\right)J_{\Delta-d/2}\left(wz_{2}\right)}{k_{h}^{2}+w^{2}}.\label{ttcb}
	\end{eqnarray}  
The integral over $z_1$ in \eqref{qiuwq} can be evaluated as follows:
\begin{eqnarray}
		&&	\int_0^{\infty} \frac{d z_1}{z_1^{d+1}} \phi_{d-1}\left(z_1, k_h\right) \phi_{d-1}\left(z_1, k_s\right) z_1 G\left(z_1, z_2, k_h+k_s, d-1\right)\nonmy\eqnmy\int_0^{\infty} dz_1\int_0^{\infty} dw\frac{2^{\frac{d}{2}-1}  {k_h^{\frac{d}{2}-1}K_{\frac{d-2}{2}}\left(k_h z_1\right)}{ z_1 \Gamma \left(\frac{d-2}{2}\right)}  
			w z_2^{d/2}	J_{\frac{d-2}{2}}\left(w z_1\right) J_{\frac{d-2}{2}}\left(w z_2\right) }{\pi 
			\left(k_h^2+w^2\right)}\nonmy\eqnmy \frac{2^{\frac{d}{2}-1} z_2^{\frac{d}{2}+1} \Gamma\left(\frac{d}{2}-1\right) k_h^{\frac{d}{2}-2} K_{\frac{d}{2}-2}\left(k_h z_2\right)}{\pi},
	\end{eqnarray}
where we used the following identities:
\begin{eqnarray}
		\int_{0}^{\infty} dz_1\; z_1 J_{\frac{d-2}{2}}\left(w z_1\right) K_{\frac{d-2}{2}}\left(k_h z_1\right)	\eqnmy\frac{w^{\frac{d}{2}-1} k_h^{1-\frac{d}{2}}}{k_h^2+w^2}\nonmy \int_{0}^{\infty} dw\;\frac{w^{d/2} J_{\frac{d-2}{2}}\left(w z_2\right)}{\left(k_h^2+w^2\right){}^2}\eqnmy\frac{1}{2} z_2 k_h^{\frac{d}{2}-2} K_{2-\frac{d}{2}}\left(k_h z_2\right)
\label{besselid}.
	\end{eqnarray} 
Plugging this into \eqref{qiuwq} and performing the integral over $z_2$ then gives
\begin{align}
				\left\langle\langle J_1\ldots J_{n-1}J_hJ_s \right \rangle \rangle_1 
=&-\int_0^{\infty}  \frac{d z_2}{z_2^{d+1}}\frac{2^{\frac{d}{2}-1}\Gamma\left(\frac{d}{2}-1\right){ z_2^{\frac{d}{2}+1}   k_h^{\frac{d}{2}-2} K_{2-\frac{d}{2}}\left(k_h z_2\right)}}{\pi} \nonumber \\ & \quad \times \ \varepsilon_s \cdot k_h \varepsilon_{h}^j \mathcal{A}_{n-1}^{\mathrm{YM}j}\left(z_2\right) \prod\limits_{a=1}^{n-1}\phi_{d-1}\left(z_2, k_a\right) \nonumber \\
=&\frac{1}{2}\mathcal{N}_{d-1} \int_0^{\infty}  \frac{d z_2}{z_2^{d+1}}\left({\frac{1}{k_h} \partial_{k_h}\phi_{d-1}\left(z_2, k_h\right)}\right)\varepsilon_s \cdot k_h \varepsilon_{h}^j \mathcal{A}_{n-1}^{\mathrm{YM}j}\left(z_2,\vec{k}\right) \prod\limits_{a=1}^{n-1}\phi_{d-1}\left(z_2, k_a\right) \nonumber \\
=&\mathcal{N}_{d-1}\frac{\varepsilon_s \cdot k_h}{2k_h} \partial_{k_h}\left\langle\langle  J_1 \ldots J_{n-1}J_h \right\rangle\rangle,
\label{aterm}
\end{align}
where $\partial_{k_h}$ acts on the hard gluon bulk-to-boundary propagator, which can be expressed in terms of a scalar bulk-to-boundary propagator dressed with a polarisation vector via \eqref{spinningprops}. To obtain the second-to-last line above we noted that
	\begin{eqnarray}
		{\frac{1}{k_h} \partial_{k_h} \phi_{d-1}\left(z_2, k_h\right)=	-\sqrt{\frac{2}{\pi }} z_2^{\frac{d}{2}+1} k_h^{\frac{d}{2}-2} K_{\frac{d-4}{2}}\left(k_h z_2\right)},
	\end{eqnarray}and we used $K_{\nu}(x) =  K_{-\nu}(x)$ for $\nu$ a multiple of $1/2$. 
	
Now let us consider the second term on right-hand-side of \eqref{ymres}. We shall denote this contribution to the soft limit by the subscript 2. Plugging this into \eqref{ev1e} then gives 
\begin{eqnarray}
	\left\langle\langle J_1\ldots J_{n-1}J_hJ_s \right \rangle \rangle_2	&=&-\int \frac{d z}{z^{d+1}} \varepsilon_s \cdot \varepsilon_h \frac{k_h \cdot \mathcal{A}^{\mathrm{YM}}_{n}(z)}{2 z k_h^2}\phi_{d-1}\left(z, k_s\right)\phi_{d-1}\left(z, k_h\right) \prod_{a=1}^{n-1} \phi_{d-1}\left(z, k_a\right) \nonmy \eqnmy
- \mathcal{N}_{d-1} \varepsilon_s \cdot \varepsilon_h \int \frac{d z}{z^{d+1}}  \frac{k_h \cdot \mathcal{A}^{\mathrm{YM}}_{n}(z)}{2  k_h^2}\phi_{d-1}\left(z, k_h\right) \prod_{a=1}^{n-1} \phi_{d-1}\left(z, k_a\right) \nonmy \eqnmy
-\mathcal{N}_{d-1} \frac{\varepsilon_s \cdot \varepsilon_h}{2 k_h^2} k_h\cdot \partial_{\varepsilon_h}\left\langle\langle J_h J_3 \ldots J_n\rangle\right\rangle 
\label{bterm}.
	\end{eqnarray}
where we used \eqref{btBs1} in the second line. Adding \eqref{aterm} to \eqref{bterm}, relabeling the soft leg as $n+1$, and summing over all class I diagrams then gives \eqref{wqkjhf}. Note that the relative minus sign in \eqref{wqkjhf} can be understood as coming from an antisymmetric structure constant which has been factored out of the color-ordered correlator.

\subsection{Gravity}

The analysis will be similar to the YM case. Let us consider the first term on the right-hand-side of \eqref{grres}. We will denote its contribution to the soft limit by the subscript 1. Plugging this into \eqref{ev1e} then gives
\begin{eqnarray}
\left\langle\langle T_1\ldots T_{n-1}T_hT_s \rangle \right \rangle_1 \eqnmy \int_0^{\infty} \frac{d z_1}{z_1^{d+1}} \frac{d z_2}{z_2^{d+1}} \phi_{d}\left(z_1, k_h\right)\phi_{d}\left(z_1, k_s\right){
			z_1^2} G_{\Delta}\left(z_1, z_2, k_h \right)  \nonmy&&\quad \times \ \left(\varepsilon_s \cdot k_h\right)^2 \varepsilon_h^i \varepsilon_h^j \mathcal{A}_{n}^{\mathrm{GR}i j} \prod\limits_{a=1}^{n-1}\phi_{d}\left(z_2, k_a\right),
\label{gr1}
\end{eqnarray}
where $\Delta=d$, the bulk-to-boundary propagators are given by \eqref{Btb}, and we expand the soft bulk-to-boundary propagator to leading order in the soft momentum
\begin{eqnarray}
		\phi_{d}\left(z,k_{s}\right)\eqnmy\mathcal{N}_{d}+\mathcal{O}\left(k_{s}\right),\,\,\,\mathcal{N}_{d}=\frac{2^{(d-1)/2}\Gamma\left(\frac{d}{2}\right)}{\sqrt{\pi}}.
\label{softbtbgr}
	\end{eqnarray} 
The scalar bulk-to-bulk propagator is given by \eqref{ttcb} with $\Delta=d$.

Using the following identities:
\begin{eqnarray}
		\int_{0}^{\infty} dz_1\; z_1 J_{\frac{d}{2}}\left(w z_1\right) K_{\frac{d}{2}}\left(k_h z_1\right)	\eqnmy\frac{w^{\frac{d}{2}} k_h^{-\frac{d}{2}}}{k_h^2+w^2}\nonmy \int_{0}^{\infty} dw\;\frac{w^{\frac{d}{2}+1} J_{\frac{d}{2}}\left(w z_2\right)}{\left(k_h^2+w^2\right){}^2}\eqnmy\frac{1}{2} z_2 k_h^{\frac{d}{2}-1} K_{\frac{d-2}{2}}\left(k_h z_2\right)\label{besselid2},
	\end{eqnarray}
we can carry out the integral over $z_1$ in \eqref{gr1} to obtain
 \begin{eqnarray}
		\left\langle\langle T_1\ldots T_{n-1}T_hT_s \rangle \right \rangle_1 \eqnmy\int_0^{\infty}  \frac{d z_2}{z_2^{d+1}}\frac{2^{\frac{d}{2}-1} \Gamma \left(\frac{d}{2}\right){
				z_2^{\frac{d}{2}+1}   k_h^{\frac{d}{2}-1}
				K_{\frac{d-2}{2}}\left(k_h z_2\right)}}{\pi }\nonmy &&\quad \times \ \left(\varepsilon_s \cdot k_h\right)^2 \varepsilon_h^i \varepsilon_h^j \mathcal{A}_{n}^{\mathrm{GR}i j} \prod\limits_{a=1}^{n-1}\phi_{d}\left(z_2, k_a\right)\nonmy\eqnmy-\mathcal{N}_d\int_0^{\infty} \frac{d z_2}{z_2^{d+1}}\frac{\left(\varepsilon_s \cdot k_h\right)^2}{2 {
				k_h}}
			 \left({
			 	\partial_{k_h}\phi_{d}\left(z_2, k_h\right)}\right)\varepsilon_h^i \varepsilon_h^j \mathcal{A}_{n}^{\mathrm{GR}i j} \prod_{a=1}^{n-1} \phi_{d}\left(z_2, k_a\right)\nonmy\eqnmy-\mathcal{N}_d\frac{(\varepsilon_s \cdot k_h)^2}{2k_h} \partial_{k_h}\left\langle \langle T_1\ldots T_{n-1}T_h\rangle\right\rangle,
\label{term1g}
	\end{eqnarray}
where the derivative acts on the bulk-to-boundary propagator of the hard graviton, which can be expressed as a scalar bulk-to-boundary propagator dressed with a polarisation as given in \eqref{spinningprops}. To obtain the third line, we used
	\begin{eqnarray}
		{
			\frac{1}{ k_h}\partial_{k_h} \phi_d\left(z_2, k_h\right)=	-\sqrt{\frac{2}{\pi }} z_2^{\frac{d}{2}+1} k_h^{\frac{d}{2}-1} K_{\frac{d-2}{2}}\left(k_h z_2\right)}.
	\end{eqnarray}

Now consider the second term on the right-hand-side of \eqref{grres}. We will denote its contribution to the soft limit with the subscript 2. Plugging this into \eqref{ev1e} then gives
	\begin{eqnarray}
		\begin{aligned}
			\left\langle\langle T_1\ldots T_{n-1}T_hT_s \rangle \right \rangle_2= & \int_0^{\infty} \frac{d z}{z^{d+1}} \phi_{d}\left(z_1, k_h\right) \phi_{d}\left(z_1, k_s\right) \frac{\varepsilon_s \cdot \varepsilon_h \varepsilon_s \cdot k_h \varepsilon_h^i k_h^j \mathcal{M}_{n}^{i j}}{k_h^2} \prod_{a=1}^{n-1} \phi_{d}\left(z_2, k_a\right)\\=&\mathcal{N}_d\frac{\varepsilon_s \cdot \varepsilon_h \varepsilon_s \cdot k_h}{2k_h^2}\int_0^{\infty} \frac{d z}{z^{d+1}} \phi_{d}\left(z_1, k_h\right) { \varepsilon_h^{(i} k_h^{j)} \mathcal{A}_{n}^{\mathrm{GR}i j}}\prod_{a=1}^{n-1} \phi_{d}\left(z_2, k_a\right)\\=&\mathcal{N}_d\frac{\varepsilon_s \cdot \varepsilon_h \varepsilon_s \cdot k_h}{2k_h^2}\varepsilon_h^{(i} k_h^{j)}\partial_{\varepsilon_{h}^{ij}}\left\langle\langle T_1\ldots T_{n-1}T_h\rangle\right\rangle,
		\end{aligned}
\label{term2g}
	\end{eqnarray}
where we used \eqref{softbtbgr} in the second line. Adding \eqref{term1g} to \eqref{term2g}, relabeling the soft leg as $n+1$, and summing over all class I diagrams then gives \eqref{grsoftresmom}.

\subsection{Further Comments on Mellin Momentum Amplitudes} \label{furthercomment}

It is interesting to note that the soft limit of class I diagrams can be expressed in a purely algebraic way in Mellin space. To see this, it is convenient to combine the soft scalar propagator with the rest of the Mellin-momentum amplitude as follows: 
\begin{equation}
\tilde{\mathcal{A}}_{n+1}=\mathcal{A}_{n+1} \phi_{\Delta}\left(z,k_{s}\right).
\end{equation} 
In the soft limit, this will rescale the Mellin momentum amplitude by a factor of $z$ in YM and will have no effect in GR. On performing the Mellin transformation of bulk-to-boundary propagators following \eqref{Btb2}, the energy derivative on the hard bulk-to-boundary propagator in \eqref{aterm} gives
\begin{eqnarray}
	\partial_{k_h} \phi_{\Delta}(s, k_h)=\frac{(\Delta-d / 2-2 s)}{k_h} \phi_{\Delta}(s, k_h) .
\end{eqnarray}
Thus, we can write \eqref{ymres} as
\begin{equation}
	\lim _{\vec{k}_{s} \rightarrow 0} \tilde{\mathcal{A}}_{n+1}^{\mathrm{YM} 1} \rightarrow \mathcal{N}_{d-1} \left(\frac{\varepsilon_{s} \cdot k_h\left(d / 2-2 s_h-1\right)}{2 k_h^2}-\frac{\varepsilon_{s} \cdot \varepsilon_h}{2 k_h^2} k_h \cdot \partial_{\varepsilon_h}\right) \mathcal{A}_n^{\mathrm{YM}}.
\end{equation}
Similarly, for gravity we find using \eqref{term1g}
\begin{eqnarray}
\lim _{\vec{k}_s \rightarrow 0} \tilde{\mathcal{A}}_{n+1}^{\text {GR1 }}\to  \mathcal{N}_d\left[-\frac{\left(\varepsilon_{s} \cdot k_h\right)^2(d/2	-2s_h)}{2 k_h^2}+\frac{\varepsilon_{s} \cdot \varepsilon_h \varepsilon_{s} \cdot k_h}{2 k_h^2} \varepsilon_h^{(i} k_h^{j)} \partial_{\varepsilon_h^{i j}}\right]\mathcal{A}_n^{\mathrm{GR}}.
	\nonmy
\end{eqnarray}

\section{Soft Limits in Momentum Space} \label{softmomentumappendix}

In this Appendix we provide some more details about the soft limit of gluon correlators by directly analyzing Witten diagrams in momentum space\footnote{CC would like to thank Savan Kharel for useful discussions on this topic.}. The following four classes of diagrams contribute to a generic YM correlator:\footnote{We use $\mathcal F$ here to distinguish expressions in momentum space from the corresponding expressions in Mellin momentum space in section \ref{sec:generalMult}.  }
\begin{figure}[H]
\centering
\subfigure[Class I]{
\scalebox{0.9}{\begin{tikzpicture}
\draw[very thick, fill = lightgray!50] (-1.5, 0) circle (0.5);
\draw[very thick, fermion, black] (-1, 0) -- (1,0);
\draw[very thick, fermion, black] (1, 0) -- ({2.5*cos(30)}, {2.5*sin(30)});
\draw[very thick, fermion, black] (1, 0) -- ({2.5*cos(-30)}, {2.5*sin(-30)});
\draw[very thick] (0, 0) circle (2.5);


\draw[very thick, black] (-1.82, 0.38) -- ({2.5*cos(160)}, {2.5*sin(160)});
\draw[very thick, black] (-1.82, -0.38) -- ({2.5*cos(200)}, {2.5*sin(200)});
\node at (-2.25, 0) {$\vdots$};

\node at (-1.5, 0) {$\mathcal{F}_n$};
\node at (1.2, 1) {$\vec k_s$};
\node at (1.2, -1) {$\vec  k_h$};
\end{tikzpicture} }

}
~ 
\subfigure[Class II]{
\scalebox{0.9}{\begin{tikzpicture}
\draw[very thick, fill = lightgray!50] (-1.5, 0) circle (0.5);
\draw[very thick, fill = lightgray!50] (1.5, 0) circle (0.5);
\draw[very thick, fermion, black] (-1, 0) -- (0,0);
\draw[very thick, fermion, black] (0, 0) -- (1,0);
\draw[very thick, fermion, black] (0, 0) -- (0,2.5);
\draw[very thick] (0, 0) circle (2.5);


\draw[very thick, black] (-1.82, 0.38) -- ({2.5*cos(160)}, {2.5*sin(160)});
\draw[very thick, black] (-1.82, -0.38) -- ({2.5*cos(200)}, {2.5*sin(200)});
\node at (-2.25, 0) {$\vdots$};


\draw[very thick, black] (1.82, 0.38) -- ({2.5*cos(20)}, {2.5*sin(20)});
\draw[very thick, black] (1.82, -0.38) -- ({2.5*cos(-20)}, {2.5*sin(-20)});
\node at (2.25, 0) {$\vdots$};

\node at (-1.5, 0) {$\mathcal{F}_L$};
\node at (1.5, 0) {$\mathcal{F}_R$};
\node at (0.3, 1) {$ \vec  k_s$};

\end{tikzpicture}} } \\
\subfigure[Class III]{
\scalebox{0.9}{\begin{tikzpicture}
\draw[very thick, fill = lightgray!50] (-1.5, 0) circle (0.5);
\draw[very thick, fermion, black] (-1, 0) -- (1,0);
\draw[very thick, fermion, black] (1, 0) -- (2.5,0);
\draw[very thick, fermion, black] (1, 0) -- ({2.5*cos(30)}, {2.5*sin(30)});
\draw[very thick, fermion, black] (1, 0) -- ({2.5*cos(-30)}, {2.5*sin(-30)});
\draw[very thick] (0, 0) circle (2.5);


\draw[very thick, black] (-1.82, 0.38) -- ({2.5*cos(160)}, {2.5*sin(160)});
\draw[very thick, black] (-1.82, -0.38) -- ({2.5*cos(200)}, {2.5*sin(200)});
\node at (-2.25, 0) {$\vdots$};

\node at (-1.5, 0) {$\mathcal{F}_n$};
\node at (2.11, -0.4) {$\vec k_{h_1}$};
\node at (1.2, 1) {$\vec k_s$};
\node at (1.2, -1) {$\vec k_{h_2}$};
\end{tikzpicture}} }
~ 
\subfigure[Class IV]{
\scalebox{0.9}{\begin{tikzpicture}
\draw[very thick, fill = lightgray!50] (-1.5, 0) circle (0.5);
\draw[very thick, fill = lightgray!50] (1.5, 0) circle (0.5);
\draw[very thick, fermion, black] (-1, 0) -- (0,0);
\draw[very thick, fermion, black] (0, 0) -- (1,0);
\draw[very thick, fermion, black] (0, 0) -- ({2.5*cos(70)}, {2.5*sin(70)});
\draw[very thick, fermion, black] (0, 0) -- ({2.5*cos(110)}, {2.5*sin(110)});
\draw[very thick] (0, 0) circle (2.5);


\draw[very thick, black] (-1.82, 0.38) -- ({2.5*cos(160)}, {2.5*sin(160)});
\draw[very thick, black] (-1.82, -0.38) -- ({2.5*cos(200)}, {2.5*sin(200)});
\node at (-2.25, 0) {$\vdots$};


\draw[very thick, black] (1.82, 0.38) -- ({2.5*cos(20)}, {2.5*sin(20)});
\draw[very thick, black] (1.82, -0.38) -- ({2.5*cos(-20)}, {2.5*sin(-20)});
\node at (2.25, 0) {$\vdots$};

\node at (-1.5, 0) {$\mathcal{F}_L$};
\node at (1.5, 0) {$\mathcal{F}_R$};
\node at ({2.25*cos(70)}, {1*sin(70)}) {$\vec k_s$};
\node at ({-2.25*cos(70)}, {1*sin(70)}) {$\vec k_h$};
\end{tikzpicture}} }
\end{figure}

A similar classification was used to prove soft theorems for superstring scattering amplitudes in flat space \cite{Sen:2017xjn}. In particular, the Weinberg soft theorem in flat space arises only from class I diagrams above while the subleading soft theorem receives contributions from classes I and II. We first consider class I diagrams, obtaining results consistent with the bootstrap procedure in section \ref{gluonsoftbootstrap}. Consider the soft limit of a general class I diagram as shown below:
\begin{eqn}
&\lim_{\vec k_s \to 0} \scalebox{0.9}{ \begin{tikzpicture}[baseline]
\draw[very thick, fill = lightgray!50] (-1.5, 0) circle (0.5);
\draw[very thick, fermion, black] (-1, 0) -- (1,0);
\draw[very thick, fermion, black] (1, 0) -- ({2.5*cos(30)}, {2.5*sin(30)});
\draw[very thick, fermion, black] (1, 0) -- ({2.5*cos(-30)}, {2.5*sin(-30)});


\draw[very thick, black] (-1.82, 0.38) -- ({2.5*cos(160)}, {2.5*sin(160)});
\draw[very thick, black] (-1.82, -0.38) -- ({2.5*cos(200)}, {2.5*sin(200)});
\node at (-2.25, 0) {$\vdots$};

\draw[very thick] (0, 0) circle (2.5);
\node at (-1.5, 0) {$\mathcal{F}_n^\mathrm{YM}$};
\node at (1.2, 1) {$\vec k_s$};
\node at (1.2, -1) {$\vec k_h$};
\node at (0, 0.5) {$\vec k_s + \vec k_h$};
\end{tikzpicture} } \equiv \lim_{\vec k_s \to 0} \mathcal F_{n+1}^{YM}\\
&= 
\lim_{\vec{k}_{s}\to0}\intsinf dz_{1}dz_{2}\mathcal{F}_n^{\mathrm{YM}, i}(z_{1})G_{ij}(z_{1},z_{2},\vec k_s+\vec k_h)e^{-(k_{s}+k_{h})z_{2}}V^{jkl}\left(-(\vec{k}_{s}+\vec{k}_{h}),\vec{k}_{s},\vec{k}_{h}\right)\varepsilon_{k}(\vec{k}_{h})\varepsilon_{l}(\vec{k}_{s}),
\end{eqn}
where the propagators and the vertex factors are defined in axial gauge \footnote{These are stated in \cite{Albayrak:2019asr}:
\begin{eqn}
G_{ij}(z_{1},z_{2},\vec{y})=\intinf\frac{d\omega}{\omega^{2}+y^{2}}\sin(\omega z_{1})\sin(\omega z_{2})\left(\delta_{ij}+\frac{y_{i}y_{j}}{\omega^{2}}\right)
\end{eqn}
\begin{eqn}\label{3vertex}
V_{jkl}\left(\vec{k}_{1},\vec{k}_{2},\vec{k}_{3}\right)=\delta_{jk}\left(\vec{k}_{1}-\vec{k}_{2}\right)_{l}+\mathrm{cyclic}.
\end{eqn}
} and the blob is defined as $\mathcal{F}_n^{\mathrm{YM}, i}(z_1)$, with $i$ a Lorentz index that contracts onto the rest of the diagram via a bulk-to-bulk propagator. Note that in the limit $\vec{k}_s \to 0$ we get
\begin{eqn}
\lim_{\vec{k}_{s}\to0} \mathcal F_{n+1}^{YM}
=\intsinf dz_{1}dz_{2}\mathcal{F}_{n}^{{\rm YM}, i}(z_{1})G_{ij}\left(z_{1},z_{2},k_{h}\right)e^{-k_{h}z_{2}}V^{jkl}\left(-\vec{k}_{h},\vec{k}_{s},\vec{k}_{h}\right)\varepsilon_{h,k}\varepsilon_{s,l}.
\end{eqn}
Performing the integral over $z_2$ then gives
\begin{eqn}
\lim_{\vec{k}_{s}\to0} \mathcal F_{n+1}^{YM}
&=\frac{\pi}{k_{h}}V^{jkl}\left(-\vec{k}_{h},\vec{k}_{s},\vec{k}_{h}\right)\varepsilon_{h,k}\varepsilon_{s,l}\\
&\times \intsinf dz_{1}\mathcal{F}_n^{\mathrm{YM}, i}(z_{1})\left[\frac{z_{1}}{2k_{h}}e^{-k_{h}z_{1}}\left(\delta_{ij}-\frac{k_{h,i}k_{h,j}}{k_{h}^{2}}\left(1+\frac{2}{k_{h}z_{1}}\right)\right)+\frac{1}{k_{h}^{2}}\frac{k_{h,i}k_{h,j}}{k_{h}^{2}}\right].
\end{eqn}
Using the vertex factor given in \eqref{3vertex} we obtain the following for the soft limit of the $(n+1)$-point function:
\begin{align}\label{YMsoft2}
\lim_{\vec{k}_{s}\to0} \mathcal  F_{n+1}^{YM}
&=\frac{1}{2k_{h}}\Bigg\{\varepsilon_{s}\cdot k_{h}\pd{}{k_{h}}\intsinf dz\varepsilon_{h,i}\mathcal{F}_n^{\mathrm{YM}, i}(z)e^{-k_{h}z}\nonumber \\
&\qquad\quad +\varepsilon_{s}\cdot\varepsilon_{h}\frac{k_{h,i}}{k_{h}}\intsinf dz \mathcal{F}_n^{\mathrm{YM}, i}(z)\left(1-e^{-k_{h}z}\right)\Bigg\},
\end{align}
where the energy derivative acts on the bulk-to-boundary propagator of the hard leg. Relabeling the soft leg as $n+1$ and summing over all class I diagrams then gives the soft limit of color-ordered YM correlator:
\begin{align}
\lim_{\vec{k}_{s}\to0}\braket{\braket{J\cdots J}}_{n+1}=&\left\{ \frac{\varepsilon_{n+1}\cdot\vec{k}_{n}}{2k_{n}}\p_{k_{n}}\braket{\braket{J\cdots J}}_{n}\right. \nonumber \\
&\left.-\frac{\varepsilon_{n+1}\cdot\varepsilon_{n}}{2k_{n}^{2}}k_{n}\cdot\p_{\varepsilon_{n}}\left(\braket{\braket{J\cdots J}}_{n}-\left.\braket{\braket{J\cdots J}}_{n}\right|_{k_{n}=0}\right)\right\} \nonumber \\
&-\left(\vec{k}_{n}\rightarrow\vec{k}_{1},\vec{\varepsilon}_{n}\rightarrow\vec{\varepsilon}_{1}\right)+...
\label{class1soft}
\end{align}
where $\left. \braket{\braket{J \cdots J}}_{n} \right|_{k_{a}=0}$ means that we set $k_a=0$ but not $\vec{k}_a$, and ... denotes the contribution of non-class I diagrams. This generalizes the result found in \eqref{eq:YMsoftOnshell4to3} to any multiplicity. However, as argued around \eqref{eq:YMsoftOnshell4to3} for $n = 3$, the last term in the equation above is a boundary contact term in position space and can therefore be neglected. This term is also absent in the bootstrap result for any multiplicity in \eqref{wqkjhf}. Note that the energy derivative in the first term in the expression above gives a double pole in the total energy $\frac{1}{k_{1...n}^2}$. This pole appears because the vertex to which the soft leg is attached is a cubic external vertex. Thus, in the soft limit we obtain a bulk-boundary propagator. This simplification is a consequence of the orthogonality of the propagators:
\begin{eqn*}
\intsinf dz\ \mbox{bulk-bulk}(z', z) \mbox{bulk-boundary}(z) \sim z' \mbox{bulk-boundary}(z')~. 
\end{eqn*}
The extra power of $z$ on the right-hand-side then leads to a higher-order energy pole in the energy.

For a diagram in class II, one can explicitly compute the soft limit in the same manner and obtain the following:
\begin{eqn}
&\lim_{\vec k_s \to 0}\begin{tikzpicture}[baseline]
\draw[very thick, fill = lightgray!50] (-1.5, 0) circle (0.5);
\draw[very thick, fill = lightgray!50] (1.5, 0) circle (0.5);
\draw[very thick, fermion, black] (-1, 0) -- (0,0);
\draw[very thick, fermion, black] (0, 0) -- (1,0);
\draw[very thick, fermion, black] (0, 0) -- (0,2.5);
\draw[very thick] (0, 0) circle (2.5);


\draw[very thick, black] (-1.82, 0.38) -- ({2.5*cos(160)}, {2.5*sin(160)});
\draw[very thick, black] (-1.82, -0.38) -- ({2.5*cos(200)}, {2.5*sin(200)});
\node at (-2.25, 0) {$\vdots$};


\draw[very thick, black] (1.82, 0.38) -- ({2.5*cos(20)}, {2.5*sin(20)});
\draw[very thick, black] (1.82, -0.38) -- ({2.5*cos(-20)}, {2.5*sin(-20)});
\node at (2.25, 0) {$\vdots$};

\node at (-1.5, 0) {$\mathcal{F}_L$};
\node at (1.5, 0) {$\mathcal{F}_R$};
\node at (-0.5, -0.25) {$\vec y_1$};
\node at (0.5, -0.25) {$\vec y_2$};
\node at (0.3, 1) {$\vec k_s$};
\end{tikzpicture} 
= 
- \frac{\vec \varepsilon_s \cdot \vec y}{y} \p_y \begin{tikzpicture}[baseline]
\draw[very thick, fill = lightgray!50] (-1.5, 0) circle (0.5);
\draw[very thick, fill = lightgray!50] (1.5, 0) circle (0.5);
\draw[very thick, fermion, black] (-1, 0) -- (1,0);
\draw[very thick] (0, 0) circle (2.5);


\draw[very thick, black] (-1.82, 0.38) -- ({2.5*cos(160)}, {2.5*sin(160)});
\draw[very thick, black] (-1.82, -0.38) -- ({2.5*cos(200)}, {2.5*sin(200)});
\node at (-2.25, 0) {$\vdots$};


\draw[very thick, black] (1.82, 0.38) -- ({2.5*cos(20)}, {2.5*sin(20)});
\draw[very thick, black] (1.82, -0.38) -- ({2.5*cos(-20)}, {2.5*sin(-20)});
\node at (2.25, 0) {$\vdots$};

\node at (-1.5, 0) {$\mathcal{F}_L$};
\node at (1.5, 0) {$\mathcal{F}_R$};
\node at (-0, -0.25) {$\vec y$};

\end{tikzpicture} \\
&+ \frac{1}{y^2} (\varepsilon_s^{i} y^j + \varepsilon_s^j y^i) \Bigg\{ \begin{tikzpicture}[baseline]
\draw[very thick, fill = lightgray!50] (-1.5, 0) circle (0.5);
\draw[very thick, fill = lightgray!50] (1.5, 0) circle (0.5);
\draw[very thick, fermion, black] (-1, 0) -- (1,0);
\draw[very thick] (0, 0) circle (2.5);


\draw[very thick, black] (-1.82, 0.38) -- ({2.5*cos(160)}, {2.5*sin(160)});
\draw[very thick, black] (-1.82, -0.38) -- ({2.5*cos(200)}, {2.5*sin(200)});
\node at (-2.25, 0) {$\vdots$};


\draw[very thick, black] (1.82, 0.38) -- ({2.5*cos(20)}, {2.5*sin(20)});
\draw[very thick, black] (1.82, -0.38) -- ({2.5*cos(-20)}, {2.5*sin(-20)});
\node at (2.25, 0) {$\vdots$};

\node at (-1.5, 0) {$\mathcal{F}_L^i$};
\node at (1.5, 0) {$\mathcal{F}_R^j$};
\node at (-0, -0.25) {$\vec y= 0$};

\end{tikzpicture} 
-  \begin{tikzpicture}[baseline]
\draw[very thick, fill = lightgray!50] (-1.5, 0) circle (0.5);
\draw[very thick, fill = lightgray!50] (1.5, 0) circle (0.5);
\draw[very thick, fermion, black] (-1, 0) -- (1,0);
\draw[very thick] (0, 0) circle (2.5);


\draw[very thick, black] (-1.82, 0.38) -- ({2.5*cos(160)}, {2.5*sin(160)});
\draw[very thick, black] (-1.82, -0.38) -- ({2.5*cos(200)}, {2.5*sin(200)});
\node at (-2.25, 0) {$\vdots$};


\draw[very thick, black] (1.82, 0.38) -- ({2.5*cos(20)}, {2.5*sin(20)});
\draw[very thick, black] (1.82, -0.38) -- ({2.5*cos(-20)}, {2.5*sin(-20)});
\node at (2.25, 0) {$\vdots$};

\node at (-1.5, 0) {$\mathcal{F}_L^i$};
\node at (1.5, 0) {$\mathcal{F}_R^j$};
\node at (-0, -0.25) {$\vec y$};
\end{tikzpicture} \Bigg\}
\end{eqn}
where $\vec{y}$ refers to the sum of boundary momenta flowing through a propagator and $y$ denotes its magnitude. In the bottom two diagrams we have displayed the free Lorentz indices on the blobs which contract with the soft polarisation and exchanged momentum.  Note that these diagrams do not exhibit energy derivatives and therefore do not give rise to double poles in the total energy as we found for class I diagrams. Moreover, class II diagrams exhibit poles in the $y$ variables, which correspond to multiparticle energy poles, and we find similar results for higher class diagrams. This is in contrast to class I diagrams which give rise to poles in the energy of individual hard legs. Hence the polarisation derivative arising from the soft limit of class I diagrams can be distinguished from the contributions of non-class I diagrams even though it does not contain a double pole in the total energy.  

\section{Soft Limit of Five-point YM} \label{5ptchecks}

In this Appendix we will verify the YM soft limit formula in \eqref{wqkjhf} at five points. Specializing this formula to $n=4$ gives
\begin{eqnarray}
\begin{aligned}
	\lim _{\vec{k}_{5} \rightarrow 0}\langle\langle J \ldots J\rangle\rangle_{5}= & \left\{\frac{\varepsilon_{5} \cdot k_4}{2 k_4} \partial_{k_4}\langle\langle J \ldots J\rangle\rangle_4-\frac{\varepsilon_{5} \cdot \varepsilon_4}{2 k_4^2} k_4 \cdot \partial_{\varepsilon_4}\langle\langle J \ldots J\rangle\rangle_4\right\} \\
	& -\left(k_4 \rightarrow k_1, \varepsilon_4 \rightarrow \varepsilon_1\right)+ \text{free of poles in }k^2_{1234},k_1, k_4.
	\label{E1c}
\end{aligned}
\end{eqnarray}
As a simple check, let us restrict our attention to terms which contain $\varepsilon_4 \cdot \varepsilon_5 \varepsilon_1 \cdot \varepsilon_3 \varepsilon_2 \cdot k_4 $. The left-hand-side of \eqref{E1c} then gives
\begin{eqnarray}
&&
\lim _{\vec{k}_{5} \rightarrow 0}\langle\langle JJJJJ\rangle\rangle|_{\varepsilon_4 \cdot \varepsilon_5 \varepsilon_1 \cdot \varepsilon_3 \varepsilon_2 \cdot k_4}
\nonmy
\eqnmy
\frac{k_{1}}{2k_{4}k_{1234}\left(k_{14}+\left|\vec{k}_{2}+\vec{k}_{3}\right|\right)\left(k_{23}+\left|\vec{k}_{2}+\vec{k}_{3}\right|\right)}+O(k_{4}^{0}).
\label{5pt1}
\end{eqnarray}
The first term on the right-hand-side of \eqref{E1c} does not contribute while the second term gives
\begin{eqnarray}
&&	-\frac{\varepsilon_{5} \cdot \varepsilon_4}{2 k_4^2} k_4 \cdot \partial_{\varepsilon_4}\langle\langle JJJJ\rangle\rangle|_{\varepsilon_4 \cdot \varepsilon_5 \varepsilon_1 \cdot \varepsilon_3 \varepsilon_2 \cdot k_4}
=
-\frac{1}{4
	k_4^2 \left(k_2+k_3+\left|\vec{k}_{2}+\vec{k}_{3}\right|\right)}
	\nonmy
	&&+\frac{k_{1}}{2k_{4}k_{1234}\left(k_{14}+\left|\vec{k}_{2}+\vec{k}_{3}\right|\right)\left(k_{23}+\left|\vec{k}_{2}+\vec{k}_{3}\right|\right)}+O(k_{4}^{0}).
		\nonmy 
\label{5pt2}
\end{eqnarray}
Taking the difference of \eqref{5pt1} and \eqref{5pt2} then gives a term which is analytic in the momenta of at least two legs and is therefore a boundary contact term. In a similar manner, we may check \eqref{E1c} more generally with the help of \texttt{MultivariateApart}\cite{Heller_2022}. In particular, we have checked that the poles in $k_{1234}^2$, $k_1$, and $k_4$ on both sides of \eqref{E1c} agree up to boundary contact terms.

We can also check the soft limit formula in Mellin-momentum space, which is convenient because this representation is free of boundary contact terms \cite{Mei:2023jkb}. We will just sketch how this works at five points and leave the details to \texttt{5ptYM.nb}. At five points, the soft limit in Mellin-momentum space is given by 
\begin{eqnarray}
	\lim _{\vec{k}_5 \rightarrow 0} \mathcal{A}_5^{\mathrm{YM}}
	\eqnmy
	-\frac{z 
		\varepsilon_5
		\cdot 
		k_4\mathcal{A}_4^{\mathrm{YM}}}{\mathcal{D}_{k_{45}}^{d-1}}+\frac{z \varepsilon_5 \cdot k_1 \mathcal{A}_4^{\mathrm{YM}}}{\mathcal{D}_{k_{15}}^{d-1}}-\varepsilon_5 \cdot \varepsilon_4\frac{k_4 \cdot \partial_{\varepsilon_4} \mathcal{A}_4^{\mathrm{YM}}}{2 z k_4^2}+\varepsilon_5 \cdot \varepsilon_1 \frac{k_1 \cdot \partial_{\varepsilon_1} \mathcal{A}_4^{\mathrm{YM}}}{2 z k_1^2}
	\nonmy
	&& 
	+ \text{free of poles in }k_4, k_1,\mathcal{D}_{k_{45}}^{d-1},\mathcal{D}_{k_{15  }}^{d-1}.
	\label{E1mmAmp}
\end{eqnarray}
Starting with an explicit formula for the five-point Mellin momentum amplitude derived in \cite{Mei:2024abu}, we compute the residues of the factorisation poles and the poles in the hard momenta and find that
\begin{eqnarray}
	\underset{\mathcal{D}_{k_{4}}^{d-1}\to 0}{\mathrm{Res}}     \lim _{\vec{k}_5 \rightarrow 0} \mathcal{A}_5^{\mathrm{YM}}
	\eqnmy
	-\frac{z
		\varepsilon_5
		\cdot 
		k_4\mathcal{A}_4^{\mathrm{YM}}}{\mathcal{D}_{k_{4}}^{d-1}},
\end{eqnarray}
\begin{eqnarray}
	\underset{\mathcal{D}_{k_{1}}^{d-1}\to 0}{\mathrm{Res}}     \lim _{\vec{k}_5 \rightarrow 0} \mathcal{A}_5^{\mathrm{YM}}
	\eqnmy
	\frac{z 
		\varepsilon_5
		\cdot 
		k_1\mathcal{A}_4^{\mathrm{YM}}}{\mathcal{D}_{k_{1}}^{d-1}},
\end{eqnarray}
\begin{eqnarray}
	\underset{k_4^2\to 0}{\mathrm{Res}}     \lim _{\vec{k}_5 \rightarrow 0} \mathcal{A}_5^{\mathrm{YM}}
	\eqnmy
	-\varepsilon_5 \cdot \varepsilon_4 \frac{k_4 \cdot \partial_{\varepsilon_4} \mathcal{A}_4^{\mathrm{YM}}}{2 z  k_4^2},
\end{eqnarray}
\begin{eqnarray}
	\underset{k_1^2\to 0}{\mathrm{Res}}     \lim _{\vec{k}_5 \rightarrow 0} \mathcal{A}_5^{\mathrm{YM}}
	\eqnmy
	\varepsilon_5 \cdot \varepsilon_1 \frac{k_1 \cdot \partial_{\varepsilon_1} \mathcal{A}_4^{\mathrm{YM}}}{2 z k_1^2},
\end{eqnarray}
in agreement with \eqref{E1mmAmp}. When checking these relations, it is important to keep track of an extra factor of $z$ which arises from the soft bulk-to-boundary scalar propagator (see the discussion in section \ref{furthercomment} for more details).

When checking the polarisation derivative terms we found that in \eqref{E1c} they come with a $1/k_h$, while the coefficients of $1/k_h^2$ are boundary contact terms, and in \eqref{E1mmAmp} they come with a $1/k_h^2$. The difference arises because when going back to momentum space the Mellin variable $s_h$ will produce $k_h$ in the numerator. One may use the Mellin delta function to write $s_h$ in terms of other Mellin variables leaving a $1/k_h^2$ pole but the two expressions will only differ by boundary contact terms \cite{Sleight:2019hfp}.

\newpage

\bibliography{references}
\bibliographystyle{JHEP}

\end{document}